# A Survey on Replica Server Placement Algorithms for Content Delivery Networks

Jagruti Sahoo, *Member, IEEE*, Mohammad A. Salahuddin, *Member, IEEE*, Roch Glitho, *Senior Member, IEEE*, Halima Elbiaze, *Member, IEEE*, and Wessam Ajib, *Member, IEEE*

*Abstract*: Content Delivery Networks (CDNs) have gained immense popularity over the years. Replica server placement is a key design issue in CDNs. It entails placing replica servers at meticulous locations, such that cost is minimized and Quality of Service (QoS) of end-users is satisfied. Many replica server placement models have been proposed in the literature of traditional CDN. As the CDN architecture is evolving through the adoption of emerging paradigms, such as, cloud computing and Network Functions Virtualization (NFV), new algorithms are being proposed. In this paper, we present a comprehensive survey of replica server placement algorithms in traditional and emerging paradigm based CDNs. We categorize the algorithms and provide a summary of their characteristics. Besides, we identify requirements for an efficient replica server placement algorithm and perform a comparison in the light of the requirements. Finally, we discuss potential avenues for further research in replica server placement in CDNs.

## I. INTRODUCTION

Content Delivery Networks (CDNs) are large distributed infrastructures of replica servers placed in strategic locations [1], [2]. By replicating content of origin server on replica servers, the content is delivered to end-users with reduced latency. CDNs support a variety of content including static content, dynamic content (on-line sports score and stock quote), streaming audio/videos, news and software downloads. Over the last few years, the emergence of over-the-top (OTT) providers (e.g. Youtube, Netflix and Hulu) has also resulted in more and more deployment of CDNs. Today, enterprises of any scale are highly dependent on CDNs to maintain and grow their business. Primarily, CDNs offer the following benefits: 1) reduced load on origin servers by offloading the delivery tasks to replica servers, 2) reduced latency by hosting content close to end-users, 3) improved content availability due to multiple delivery points, and 4) reduced overhead on network

Manuscript submitted on March 16, 2016.
J. Sahoo is with South Carolina State University, Orangeburg, SC, USA. (e-mail: jsahoo@scsu.edu)
R. Glitho is with Concordia University, Montréal, H3G 1M8, Canada and University of Western Cape, Bellville 7535, South Africa (e-mail: glitho@ece.concordia.ca)
M. A. Salahuddin, H. Elbiaze, and W. Ajib are with Université du Québec à Montréal, Montréal, Canada (e-mail: mohammad.salahuddin@ieee.org, elbiaze.halima@uqam.ca, ajib.wessam@uqam.ca)

backbone and alleviated congestion by avoiding long distance transmission of voluminous traffic such as videos.

CDNs where the replica servers are built using conventional web technologies are referred to as traditional CDNs. The replica servers are either dedicated servers or denote a dedicated storage space in a shared infrastructure. Examples of traditional CDNs include Akamai [3], Limelight [4] and Level 3 [5]. Lately, CDNs have been architected using emerging paradigms such as cloud computing and Network Functions Virtualization (NFV). In the next subsections, we discuss CDNs based on emerging paradigms, replica server placement in CDNs, existing surveys in CDNs, our contributions and organization of the paper.

### A. CDNs based on Emerging Paradigms

According to Cisco Visual Networking Index (VNI) 2015, the global IP traffic has increased fivefold over the past five years, and will increase threefold over the next five years [6]. The unprecedented content growth can only be tackled by traditional CDN having the required scalability. Moreover, under or over-provisioning of resources in replica servers reduce the resource allocation efficiency and increase the cost of traditional CDNs. The growing popularity of videos also results in growing end-user expectations and increases the QoS (e.g. latency) burden on traditional CDNs.

In order to improve scalability, cost efficiency, resource efficiency and QoS, CDNs can be designed by embracing emerging paradigms such as cloud computing [7], [8], [9] and NFV [10]. Cloud computing [11], [12] has several inherent advantages, such as, scalability, on-demand resource allocation, flexible pricing model (pay-as-you-go), and easy applications and services provisioning. CDNs can leverage cloud computing; the new paradigm is referred to as cloud based CDN. In cloud based CDN, replica servers are provisioned as cloud applications on top of Infrastructure as a Service (IaaS) or provisioned using Platform as a Service (PaaS) of cloud computing. Cloud based CDNs can also offer value added services using Software as a Service (SaaS) applications which are provisioned using Platform as a Service (PaaS). SaaS based value added services provided to the content providers include analytics and Digital Rights Management (DRM). Similarly, Social TV is one such SaaS based value added service provided to the end-users. Commercial cloud based CDNs include Rackspace [13], Amazon CloudFront [14], and CloudFlare [15]. In the literature, many architectures have also been proposed for cloud based CDNs. Some of them are ActiveCDN [16], MetaCDN [17], MediawiseMCCO [18] and CoDaaS [19].

NFV involves implementing the network equipment and functionalities in pieces of software, called Virtual Network Functions (VNF) that can run on commodity server hardware and can be moved or instantiated in various locations in the network, without the need for equipment installation. Using NFV, a CDN can be designed as a collection of loosely coupled virtual functionalities (i.e., VNFs), which leads to a substantial reduction in equipment cost and also the time needed to install or even expand the CDN infrastructure. It offers all the advantages offered by cloud computing. Use cases of NFV based CDNs have been identified by ETSI [11]. Apart from this, few NFV based CDN architectures have been proposed in the literature [20], [21], [22].

When designing CDN architecture, the appropriate emerging paradigm can be selected based on the requirement of content delivery scenarios in CDN. Moreover, one paradigm can bring some advantages over the other. For example, through dynamic service chaining, NFV can bring more service agility to CDN in creating and deploying new, innovative and highly customized services compared to cloud based CDNs. This advantage of NFV paradigm allows the CDN provider to enjoy improved time-to-market and a larger market share. Thus, NFV based CDN is more suitable than cloud based CDN in scenarios such as value-added service delivery, video optimization and scenarios that involve real-time service requirements from end-users. In these scenarios, NFV instantiates and chains VNFs on-demand based on end-user's requirements (e.g. specific type of video processing functionality based on end-user device). In cloud based CDNs, the value-added services are delivered as SaaS which is less agile compared to VNFs. Moreover, in case of NFV, the locations for deploying VNFs is flexible i.e. VNFs can be instantiated on network elements such as routers, switches that has NFV capability, allowing flexibility to CDN provider in adding or removing the instances of CDN control functionality (e.g. request router) based on the need (e.g. increase or decrease in number of end-user requests). On the contrary, cloud based deployment of such CDN entities may introduce some delay. However, Cloud based CDN is still preferred in scenarios that is less dynamic and has less stringent delay requirements.

### B. Replica Server Placement in CDNs

The overall efficiency of CDN is achieved when CDN is able to deliver content with high performance (quantified as reduced latency or strict bound on QoS) and low cost. Achieving these two goals influences the key design problems in CDNs: replica server placement and content placement. Replica server placement involves meticulous selection of locations to place the replica servers; whereas, content placement refers to the selection of appropriate replica servers to host a given set of content objects. The basic difference between replica server placement and content placement is that in the former, new locations are determined to place replica servers, where the replica servers contains a full or a partial replica. In content placement, a set of content objects is distributed among a subset of the deployed replica servers, referred to as replication strategy. The controller placement problem in Software-Defined-Networking [23] based network architectures is related to replica server placement as the SDN controllers are also placed strategically in suitable geographic locations.

In this paper, we focus on replica server placement algorithms proposed for traditional CDNs and emerging paradigm based CDNs. Generally, the replica server placement problem is defined as follows. Given a set of candidate locations of replica servers, a set of end-user locations, the replica server placement involves finding the optimal number and location of replica servers from the candidate set so that each end-user must be assigned to one of the replica servers and the cost of CDN provider is minimized. There can be constraints on various parameters, such as, capacity of replica servers, bandwidth capacity, QoS requirement of end-users, etc. The CDN provider executes a suitable algorithm to solve the replica server placement problem. Generally, replica server placement problem is NP-Hard. Hence, many approximation algorithms and heuristics have been proposed in the literature (traditional CDNs [24] [25] [26] [27], cloud based CDN [28] [29] [30] and NFV based CDN [31]. In cloud based CDN, the replica servers are virtual replica servers as they are built on leased virtualized resources. Similarly, in NFV based CDN, the replica server placement entails placement of virtual CDN nodes (Cache nodes, transcoders, etc.) implemented as VNFs, a problem referred to as VNF placement.

Traditional CDNs and emerging paradigm based CDNs focus on different aspects while placing the replica servers. Some of the aspects are discussed as follows. First, Replica server placement algorithms in traditional CDN do not consider real-time end-user demand and are primarily of offline[1] in nature except some works such as [33] [34] that place the requested content when the end-user request arrives; whereas Cloud based CDN and NFV based CDN has the ability to support offline as well as online[2] placement of replica servers. In Cloud based CDNs, the online algorithms are used to address optimal resource allocation such as finding the appropriate server to lease at a given time instant to accommodate spatial and temporal variations in end-user demands [92] and to find optimal amount of leased resources to meet QoS [89]. Second, some cloud providers exhibit dynamic behavior e.g. they vary their resource prices when electricity cost varies in regional power systems. Hence, in cloud based CDN, the dynamic resource price is considered in finding the placement of replica servers [88]; whereas traditional CDN algorithms do not have such considerations. Third, most traditional CDN algorithms [68] [70] specify update strategies to ensure consistency. However, emerging paradigm based CDNs do not focus on update methods except very few [29] [30]. Fourth, with NFV based CDN, the placement algorithms [98] consider different service requirements (i.e. functionalities such as Firewall, Video Optimizer) of end-users that require a strict enforcement on the order in which VNFs are placed; whereas such aspects are not seen in traditional CDNs.

---

[1] An offline replica server placement algorithm is one that executes by considering predicted end-user requests over a long period. Unlike online algorithms, offline algorithms do not respond to end-user requests.

[2] An online replica server placement algorithm is one that executes when an end-user sends request to access content delivery services. Algorithms that execute based on end-user requests at very small time intervals are also considered as online.

### C. Existing Surveys in CDNs

Up-to-date, no substantial effort has been made towards discussing the literature on replica server placement algorithms in CDNs. Bartolini *et al.* [35] and Pathan *et al.* [36] survey architectures and algorithms for traditional CDNs. However, they lack exhaustive discussion on server placement algorithms. Al-Sheyeji *et al.* [37] proposes an evaluation framework for comparing replica server placement algorithms. However, they cover only a handful of algorithms from traditional CDN. Wang *et al.* [38] present an exhaustive survey of Cloud based CDNs. But, their survey focuses only on architectures and does not discuss the optimization models and algorithms that have been proposed for cloud based CDNs. Fu *et al.* [39] present a general survey on QoS aware replica server placement algorithms that lacks discussion from the CDN perspective. In [40], the authors present a survey on resource management and scheduling algorithms for cloud mobile media networks. Regarding placement of replica servers in cloud based CDNs, only one work (i.e. Chen *et al.* [29]) is discussed in [40].

### D. Our Contributions

In this paper, we present a comprehensive survey of replica server placement algorithms in traditional CDNs and in emerging paradigm based CDNs. We discuss the theoretical problems used to model replica server placement before providing an in-depth and thorough discussion of the algorithms for replica server placement. Moreover, we propose a taxonomy for the replica server placement algorithms in traditional CDN and provide a summary for algorithms in each category of the taxonomy. We also discuss replica server placement algorithms proposed for cloud based CDN and NFV based CDN. Furthermore, we identify a set of requirements that are expected from an efficient replica server placement algorithm and conduct an evaluation of the algorithms. The evaluation can serve as an important tool in selecting an algorithm based on relative preference of requirements. We identify potential research issues related to replica server placement in CDNs and we provide insights on possible approaches to overcome those issues.

### E. Paper Organization

The reminder of the paper is organized as follows. Section II provides an overview of traditional CDN and emerging paradigms based CDN. Section III presents a background on replica server placement in CDN. Section IV depicts the theoretical problems on replica server placement in CDN. In Section V and Section VI, we discuss the replica server placement algorithms for traditional CDN and emerging paradigms based CDN respectively. Section VII presents the comparison of the algorithms. In Section VIII, we highlight future research directions for replica server placement. Finally, we conclude the paper in Section IX.

## II. OVERVIEW OF TRADITIONAL CDN AND EMERGING PARADIGMS BASED CDN

In this section, we discuss existing business model and basic mechanisms for traditional CDN and CDN based on cloud computing paradigm. Since literature on NFV based CDN lacks business model, we discuss a potential use case of NFV based CDN. We also present general view of each type of CDN.

### A. Traditional CDN

According to [2], the main business actors in traditional CDN include end-user, content provider and CDN provider. The end-user is the entity that consumes the content (e.g. video) from content provider. The content provider (e.g. YouTube) is the entity that owns the content or obtained the rights sell the content. The CDN provider (e.g. Akamai) is the entity that owns replica servers in strategic locations and offers content delivery services to the content providers. Fig. 1 shows the business model for traditional CDN. As shown in Fig. 1, business relationship exists between content provider and CDN provider in which the CDN provider gets paid by the content provider for hosting its content. Apart from the above business actors, a traditional CDN business model also involves entity such as Internet Service Provider (ISP). CDN provider can have business contract with ISP to rent their infrastructure (e.g. servers and datacenters) to deploy replica servers.

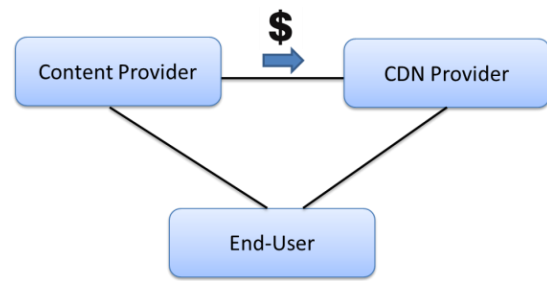

Fig. 1 Traditional CDN Business Model (Dollar with arrow sign shows the revenue stream)

Generally, a traditional CDN architecture consists of the following components: origin server, replica server and end-users. The origin server is the content provider's main server and holds the actual content. Replica servers are owned and managed by CDN providers and cache copies of the content. CDN operations can be categorized into three key phases; 1) Content Distribution, in which the contents of the origin server are replicated on replica servers; 2) Request Routing, in which end-users requests are redirected to suitable replica servers; and 3) Content Delivery, in which the content is retrieved from replica servers and sent to the end-users. Fig. 2 shows a general view of traditional CDN with the depiction of above three phases.

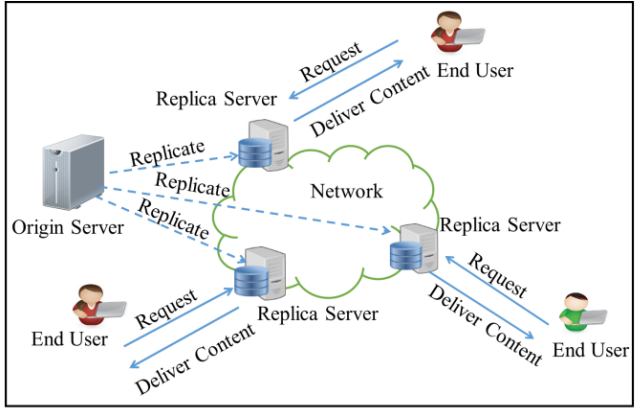

Fig. 2. General view of a traditional CDN architecture (The underlying network can be a physical or logical network)

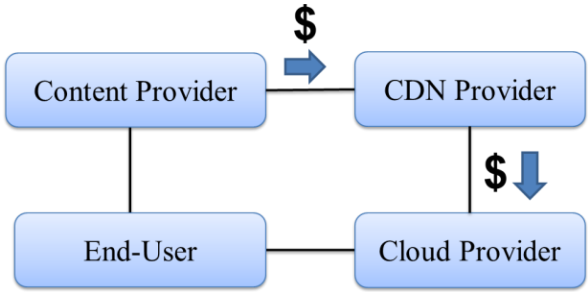

Fig. 3. Cloud based CDN Business Model (Dollar with arrow sign shows the revenue stream)

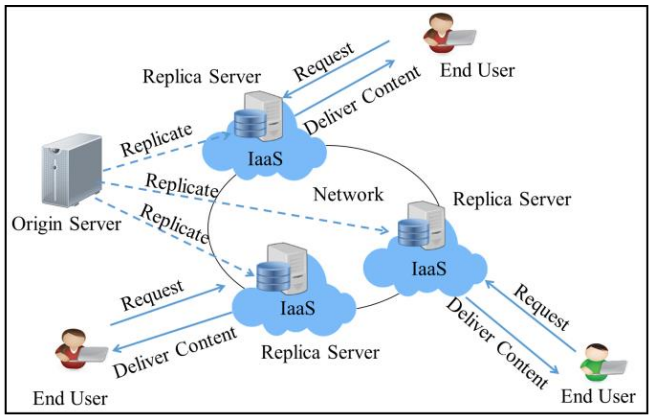

Fig. 4. General view of Cloud based CDN (The underlying network can be a physical or logical network)

### B. CDN based on Emerging Paradigms

*1) Cloud-based CDN:* A cloud based CDN includes the following business actors: end-user, content provider, CDN provider and cloud provider [17]. The cloud provider owns the cloud infrastructure in various locations. In order to deploy replica servers, the CDN provider simply leases the resources from one more cloud providers using a pay-as-you-go pricing model. In addition, the cloud provider sometimes offers a PaaS to CDN provider for the development and management of value-added services. Fig. 3 shows the business model, where CDN provider pays cloud provider for the leased resources and content provider pays CDN provider for the content delivery services.

However, like in many business models, the same actor can play several roles. Similarly, a given business actor can provide both CDN and cloud services i.e. the CDN provider deploys replica servers in its own private cloud. Fig. 4 shows a general view of Cloud based CDN. The three key phases of CDN operations in traditional CDN architecture are also performed in Cloud based CDN. In the cloud based CDN architecture proposed in [30], the content provider plays the role of CDN provider and is referred to as CCDN provider. The CCDN provider deploys its own cloud based CDN by leasing IaaS resources from cloud providers (e.g. Amazon EC2) and transit network providers that use virtual networking technology to provide a network overlay between different cloud providers. The architecture supports two methods of negotiation with cloud providers and transit network providers: direct and through some brokerage service. To place replica servers, the CCDN provider first identifies potential end-users in a geographically defined service area. The service area grid is then partitioned into a number of smaller service areas, known as clusters. Each cluster is assigned to a cloud provider. The replica servers are then executed and placed on cloud sites of the cloud provider considering the requirements of end-users located in the service cluster.

*2) NFV based CDN*

NFV use cases for virtualizing CDN entities such as cache nodes, request routers, were first specified by European Telecommunications Standards Institute (ETSI) [11]. One recent development in NFV based CDN show use of NFV paradigm for multimedia delivery where transcoding functionality is virtualized in addition to the CDN entities [41]. Because of dynamic chaining of VNFs, NFV allows a rapid introduction of novel multimedia based value-based services. Fig. 5 shows a potential use case of NFV based CDN where value-added services are deployed on-demand. The use case entails delivering video advertisements to end-users. This use case involves virtualization of the video processing functionalities of the replica servers, such as, transcoding, video mixing and video compressing. Considering that replica servers act as NFVI nodes, the virtual functionalities are instantiated and chained in a dynamic way, thereby allowing end-users to access a variety of novel value added services.

A NFV based CDN architecture may contain an NFV management and orchestration [41] that can be used to manage the instantiation and chaining of VNFs. Moreover, it manages the entire life cycle of VNFs. Dynamic chaining will take place by using an SDN controller. Other paradigms such as active networks can be investigated to perform dynamic chaining of VNFs in NFV based CDN.

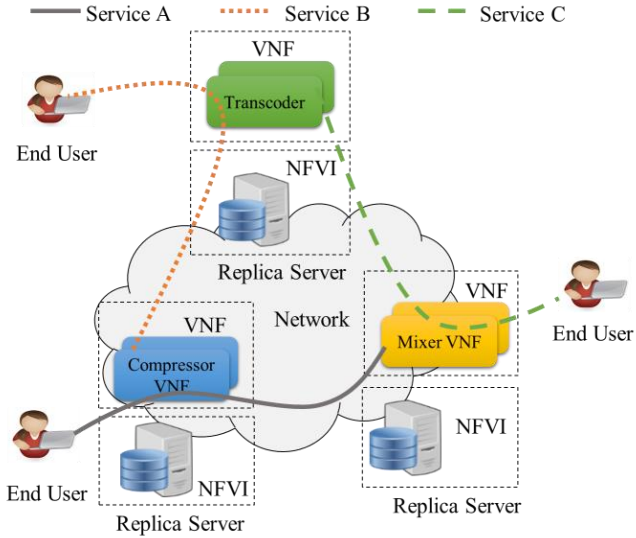

Fig. 5. NFV based CDN Use case ((The underlying network can be a physical or logical network)

## III. PARAMETERS FOR REPLICA SERVER PLACEMENT IN CDN

The replica server placement problem is formulated as a constrained or unconstrained optimization problem. The objectives and constraints are designed based on one or more input parameters. Primarily, cost of the CDN provider is considered as the objective function; while optimization of QoS has also been noticed in some cases. Constraints typically include QoS requirements of end-users such as latency, replica server capacity (e.g. compute, storage and bandwidth), maximum number of replica servers, budget of server placement, etc. We discuss the input parameters in two main categories: 1) Cost related parameters, and 2) Network related parameters, as follows.

### A. Cost Related Parameters

The cost of CDN provider basically includes the cost of deploying replica servers and a network cost in delivering content to end-users and/or updating content on replica servers. Accordingly, we define the following types of costs. 1) *Deployment cost* which is the cost of deploying replica server through buying/leasing servers or resources (compute, storage and bandwidth). It is calculated using the unit resource cost and the size of replica to be placed at the server. 2) *Delivery cost*, which is the network cost incurred in delivering content from replica server (or origin server) to the end-users. It is basically calculated using the unit bandwidth cost and the end-user demand which are provided as input to the algorithm. 3) *Update cost*, which is the network cost incurred in disseminating the updated content from origin server to replica servers or from one replica server to a group of replica servers (e.g. when an end-user performs a write operation [42]). It is calculated using the unit bandwidth cost and an update parameter that denotes the rate at which the replica servers are refreshed.

### B. Network Related Parameters

Replica server placement algorithms consider a network made up of candidate locations of replica servers (e.g. data centers, network elements), location of end-users and the communication links that connect them. However, the network is modelled as different topologies by arranging nodes in a particular order. Given a network of nodes, the link between any two nodes is associated with several performance metrics that denote the QoS metric of that link. Network related parameters are discussed as follows.

*1) Network Topology:* The replica server placement algorithms proposed for CDNs mainly use a flat [53] [54] and/or a hierarchical topology [44] [55]. A flat network topology is the one in which the all the nodes i.e. the end-users, the origin server as well as the potential sites (routers and datacenters) are at the same level. A flat network topology can have a full mesh or partial-mesh structure. Replica server placement algorithms proposed in [24] [26] [28] and [31]]) determine the placement in a flat network topology. On the other hand, a hierarchical network topology contains multiple levels and the nodes are placed at different levels. A typical hierarchical network topology adopted by many replica server placement algorithms in traditional CDNs is a tree in which the origin server is placed at the root, the end-users are located at the last level and the potential sites are placed at the intermediate levels [42] [44] [55] [56]. Another example is a four layer hierarchical network topology used by a datacenter placement algorithm [41], where each layer contains a group of potential sites (i.e. ISP's Point-of-Presence nodes) that are connected by either star, full mesh or partial mesh topology [41].

*2) Network Performance Metrics*: These are the parameters that provide a measure of the performance of the link between replica server and end-user in the network topology. The network performance parameters are either included in the objective function or considered as constraints. In this regard, latency [43] , hop-count [44], available bandwidth [45] and link quality [46] are the network parameters considered by the replica server placement algorithms proposed for CDNs.

*a) Latency*: Replica server placement schemes such as the ones proposed in [24] [47] [48] consider latency as an input parameter. To find the placement of replica server, the latency needs to be obtained between every pair of replica server and end-user in the network topology. Primarily, latency in replica server placement refers to the time required by content to be sent from a replica server to the end-user. Latency can be obtained by measuring the Round-Trip Time (RTT). Latency can also be modelled as a function of geographical distance between the nodes [49]. The latter approach is useful when the RTT information is not available. Latency is either used in the objective function, where it is either minimized or it is used as a constraint to satisfy the latency requirements of end-users.

Some replica server placement algorithms rely on popular technique such as Global Network Positioning (GNP) [50] [26], in which latency can be computed in a scalable and timely manner. GNP is coordinate-based technique that models the Internet latency in a multi-dimensional space. The latency between any pair of nodes is approximated by the Euclidean distance between their multi-dimensional coordinates. The coordinates in GNP technique also serve as building blocks to design replica server placement heuristics [25].

*b) Hop-Count*: Some replica server placement algorithms [44] [46] consider the hop-count between the replica server and end-user as an input parameter and use it as the metric to select the location of replica servers. It is defined as the number of hops in the shortest routing path between the replica server and end-user. Hop-count is used in the objective function considering the fact that minimizing the number of hops will reduce the latency.

*c) Available Bandwidth:* The available bandwidth is used as QoS parameter in replica server placement algorithms such as [45] that consider Autonomous Systems (ASs) as the potential sites to place replica servers. In [45], available bandwidth denotes the minimum bandwidth of the link between two ASs. This information can be obtained from ISP or estimated by using the methods proposed in ( [51], [52]).

## IV. THEORETICAL MODELS OF REPLICA SERVER PLACEMENT IN CDN

The replica server placement problem in CDN has been modelled after well-known theoretical models, such as, *facility location*, *connected facility location*, *K-median*, *minimum K-center*, *K-cache location*. In the following sub-sections, we discuss these placement models in detail along with the corresponding Integer Linear Programming (ILP) formulation wherever applicable. The network topology used in the ILPs is modelled as a graph $G = (V, E)$, where $V$ is the set of nodes, where each node represents a potential replica server location or end-user location[3] and $E$ is the set of edges. Let $F$ denote the set of potential locations of replica servers and $D$ denote the set of end-users. Also, we denote $M$ as the number of potential location of replica servers and $N$ as the number of end-users, that is, $|F|=M$ and $|D|=N$, respectively. Placing a replica server at location $i \in F$ incurs a cost of $f_i$. Network cost $c_{ij}$ represents the cost to deliver $d_j$ units of demand[4] from replica server location $i$ to an end-user $j$. TABLE I shows the meaning of decision variables used in this section.

TABLE I
DECISION VARIABLES

| Variables | Meaning |
|---|---|
| $x_{ij}$ | Binary variable $x_{ij} = 1$ if end-user $j$ is assigned to location $i$ |
| $y_i$ | Binary variable $y_i = 1$ if a replica server (or cache server) is placed on location $i$ |
| $z_{ik}$ | Binary variable $z_{ik} = 1$ if the edge between location $i$ and $k$ is included in the minimum cost tree. |
| Q | Integer variable that denotes the maximum delivery cost. |

### A. Facility Location

The facility location model [57] involves opening facilities with lower cost in order to provide service to one or more cities. The cost includes the opening cost of facilities and the delivery cost of delivering service from the facilities to the cities. When the replica server placement problem is modelled after facility location model, the facilities represent the replica servers and the cities represent the end-users. The objective is to find an optimal subset of the replica server locations in $F$ and connect every end-user in $D$ to one of the replica servers, such that, the total cost i.e. opening cost and delivery cost is minimized. There are variants of the facility location model based on the server capacity and distributing end-user load. There exist two variants based on the server capacity: *Capacitated* and *Uncapacitated*. Capacitated variant [27] of the facility location model includes a constraint on the server capacity requiring that each replica server can serve end-user requests within the capacity available at that location. The server capacities are very crucial constraints violations of which can disrupt a large number of connections and should be avoided. The capacitated facility location model has again two variants: *soft capacitated and hard capacitated*. In the soft capacitated version, multiple replica servers with different capacities can be placed in any potential location. On the other hand, in the hard capacitated version, at most one replica server is allowed to be placed per location. Uncapacitated variant [58] does not constrain the server capacity, thus allowing replica servers to serve any number of end-user requests.

The facility location model can be further subdivided into *splittable* and *unsplittable* based on whether end-user load is distributed among replica servers. In the splittable version, the end-user load is split among the replica servers; whereas, in the unsplittable version end-user is served by exactly one replica server. The replica server placement problem studied in [27] is modelled after splittable hard-capacitated facility location model. The unsplittable hard-capacitated facility location model is analyzed in [59].

An ILP formulation of general facility location model (i.e. unsplittable uncapacitated) is given as follows.

---

[3] We refer to end-user location simply as end-user in the rest of this paper.

[4] The terms "Demand" and "workload; are used interchangeably in the paper.

$$\text{Min } C = \sum_{i \in F} f_i y_i + \sum_{i \in F} \sum_{j \in D} d_j c_{ij} x_{ij} \qquad (1)$$

Subject to

$$y_i \geq x_{ij} \ \forall i \in F, j \in D \qquad (2)$$

$$\sum_{i \in F} x_{ij} = 1 \ \forall j \in D \qquad (3)$$

$$y_i \in \{0,1\} \qquad (4)$$

$$x_{ij} \in \{0,1\} \qquad (5)$$

In Eq(1), the first term denotes the opening cost (i.e. deployment cost as defined in Section III.A and the second term denotes the delivery cost. Constraint (2) indicates that an end-user $j$ is assigned to location $i$ only if a replica server has been placed on $i$. Constraint (3) ensures that an end-user must be assigned to exactly one of the replica server locations. Constraint (4) and constraint (5) indicate the domain restrictions for the decision variables $y_i$ and $x_{ij}$ respectively.

### B. Connected Facility Location

The update of replica servers is a typical phenomenon in CDNs that host dynamic content (stock quote updates) or interactive applications (online social networks). In order to provide precise and fresh content to the end-users, the replica servers need to be consistent and synchronized with each other. This is achieved by distributing the update to all the replica servers that hold a copy of the content. The update is sent by the origin server or any replica server at which the content has changed. However, the update cost increases with an increase in replica servers, the network distance between them and the update rate. In case of interactive applications, the update rate is the number of write requests initiated by the end-users; whereas, in case of dynamic content, the update rate is the frequency with which the content changes at the origin server.

As evident, the placement of replica servers will not be cost-effective if the update cost is neglected. The replica server placement problem with update awareness is generally modelled after the *connected facility location* model [28], which involves placing replica servers at optimal locations, optimally assigning end-users to the replica servers and connecting the replica servers through an optimal distribution topology. This differs from the general facility location model by imposing a connectivity requirement among the replica servers and origin server, which is motivated by the requirements of the CDN applications (e.g. interactive applications and dynamic content). The cost function in the connected facility location model is obtained by introducing an update cost in Eq. (1). The update rate is also an extra parameter that controls the placement. It is assumed that the replica servers will be interconnected through a minimum cost spanning tree. Thus, the update cost is the cost of that tree (e.g. Steiner tree [60]) scaled by the update parameter. The objective function in the ILP formulation (including constraints (2)-(5)) of the connected facility location model is given as follows.

$$\text{Min } C = \sum_{i \in F} f_i y_i + \sum_{i \in F} \sum_{j \in D} d_j c_{ij} x_{ij} + \alpha \sum_{i \in F} \sum_{k \in F, i \neq k} w_{ik} z_{ik} \qquad (6)$$

$$z_{ik} \in \{0,1\} \qquad (7)$$

Where the third term in Eq (6) denotes the update cost. The parameter $\alpha$ is the update parameter and $w_{ik}$ is the amount of content transferred from the replica server $i$ to replica server $k$ during an update procedure. Note that the update cost in Eq. (6) corresponds to cost of tree that connects all replica servers. With a slight modification of the update cost, a tree that connects the origin server to all replica servers can be obtained. Constraint (7) denotes the domain restriction for the decision variable $z_{ik}$.

One variant of the connected facility location model i.e. soft capacitated connected facility location model has been studied in CDN [28].

### C. K-median

In the *K-median* model, the objective is to select $K$ replica server locations in order to minimize the cost. The main difference with facility location model is that the $K$-median model does not involve costs of placing a replica server. Instead, the number of replica servers to be placed at a location is fixed or bounded by a budget i.e. $K$, which is specified as an input. ILP formulation of the $K$-median model including constraint (2)-(5) is as follows.

$$\text{Min } C = \sum_{i \in F} \sum_{j \in D} d_j c_{ij} x_{ij} \qquad (8)$$

$$\sum_{i \in F} y_i \leq K \qquad (9)$$

The objective function in Eq (8) denotes the delivery cost. Constraint (9) ensures that the number of replica servers is bounded by $K$.

Uncapacitated version of the $K$-median model has been studied in CDN [58] [61]. One generalization of the $K$-median model involves the opening of replica servers of $T$ different types, where the number of replica servers of type $i$, $i$=1,2,...$T$ is bounded by $k_i$. The $K$-median model is then a special case ($T$=1) of this generalization, where all replica servers are of the same type. When $T$=2, the problem is studied as red-blue median problem in CDN [62].

### D. Minimum K-Center

In the *minimum K-center* model [63] [64] $K$ replica servers are placed so that the maximum delivery cost between an end-user and its nearest replica server is minimized. *Number of Centers* is another model similar to minimum $K$-center model. It involves finding a smallest set of replica server locations subject to a constraint on the delivery cost between any end-user and the nearest replica server location. The ILP formulation for the minimum K-Center problem is given as follows (constraints (2)-(5) are included)

$$\text{Min } C = Q \qquad (10)$$

$$\sum_{i \in F} y_i = K \tag{11}$$

$$\sum_{i \in F} c_{ij} x_{ij} \leq Q, \forall j \in D \tag{12}$$

The objective function in Eq (10) denotes the maximum delivery cost. Constraint (11) ensures that the number of replica servers is $K$. Constraint (12) ensures that $Q$ is the maximum delivery cost between any end-user and the location to which it is assigned.

### E.  K-Cache Location

The $K$-cache location model [65] involves finding the optimal location of $K$ cache servers that minimizes the total delivery cost. It is similar to the $K$-median model with the exception that each cache server is associated with a hit ratio for each end-user. If the request from an end-user is not satisfied at the cache server, then an extra cost is incurred by transmitting the requested content from the origin server to the cache server. The ILP formulation (including constraints (2)-(5) and (9)) of the $K$-cache location model is given as follows.

$$C = \sum_{i \in F} \sum_{j \in D} d_j \Big[ h_j c_{ij} + (1 - h_j)(\delta_i + c_{ij}) \Big] x_{ij} \tag{10}$$

where $h_j$ is the hit ratio associated with content requested by end user $j$ and $\delta_i$ is the network cost incurred in transmitting content from origin server to cache server $i$.

## V.  REPLICA SERVER PLACEMENT ALGORITHMS FOR TRADITIONAL CDN

Replica server placement algorithms for traditional CDNs can be broadly classified into the following categories (See Fig. 6): 1) QoS aware, 2) Consistency aware, 3) Energy Aware, and ) Others.

### A.  QoS Aware

QoS aware algorithms either optimize one more QoS parameters or provide guarantee on the QoS parameters. Thus, they can be divided into the following two categories: 1) Optimized QoS and, 2) Bounded QoS.

*1) Optimized QoS:* In [24][63], optimized QoS based replica server placement algorithms are proposed. The replica server placement problem is modelled as a slight variation of the minimum $K$-center problem, where the QoS parameter (latency) is considered as the cost function to be optimized. RTT is used as the latency indicator.

The authors in [24] discuss $l$-greedy algorithm and Transit Node Heuristic to solve the minimum $K$-center problem. The $l$-greedy algorithm allows for $l$-steps backtracking i.e. $l$ number of already placed replica servers are replaced with $l+1$ number of new replica servers that will provide lowest latency among all combinations of $l+1$ new replica servers. The backtracking continues until $K$ replica servers are chosen. The second algorithm Transit Node Heuristic chooses top $K$ nodes sorted according to their node degrees[5]. The motivation for this heuristic is based on the assumption that the nodes having higher node degree can connect to more nodes (i.e. end-users) with lower latency and hence are the best candidates for hosting replica servers.

In [43] and [44], algorithms are proposed to address placement of streaming servers with optimized QoS. The placement involves optimally deciding the number and location of streaming servers to deliver the streaming content to end-users, such that the QoS is minimized. The streaming content is delivered through a multicast or broadcast protocol. In case of multicast, content distribution trees are built, which are rooted at the streaming servers and end-users are considered as the leaves of the trees. Bhulai *et al.* [43] addresses the placement of streaming servers in CDN by considering two delivery protocols: Periodic Broadcast (PB) protocol and Hierarchical Multicast Stream Merging (HMSM) protocol. For PB protocol, the QoS metric is computed based on hop-count of the edges in the delivery trees. For HMSM protocol, the aggregate requests from all end-users are considered along with the hop-count. The placement of streaming servers is modelled using Mixed Integer Programming (MIP). Greedy and cluster-based heuristics are proposed to find suitable placement. Once the streaming servers are placed, end-users are assigned to their nearest streaming server and delivery trees are built using the shortest path routing. In [44], the replica server placement problem involves finding optimal location of a given number of replica servers in a content distribution tree. The QoS metric in the objective includes the weighted hop-count between an end-user and its nearest replica server, where the weight denotes the aggregated demand originated at the end-user. A Genetic algorithm (GA) is developed to find the placement. For a small number of replica servers, the greedy algorithm has the same cost function value as the GA. However as the number of replica server increases, the GA outperforms the greedy algorithm by achieving a significantly lower cost.

Through improved $K$-means clustering algorithm Yin *et al.* [26] solve the media server placement problem which is modelled as *uncapacitated facility location* model. Here, all end-users are clustered into $K$ clusters and the centroids are updated by repeating the clustering procedure. Finally, for each cluster, a set of points with the lowest deployment costs are selected. Then, the point that yields the minimum latency (i.e. closest to the centroid of a cluster) is selected as the replica server. The algorithm optimizes the trade-off between the deployment cost and the latency. One important feature of this algorithm is that all the end-user locations are considered as potential locations for replica servers and hence ensures the scalability of the algorithm.

---

[5] Node degree or fanout of a node is defined as the number of directly connected neighbors (i.e. nodes that are reachable in one hop) of that node.

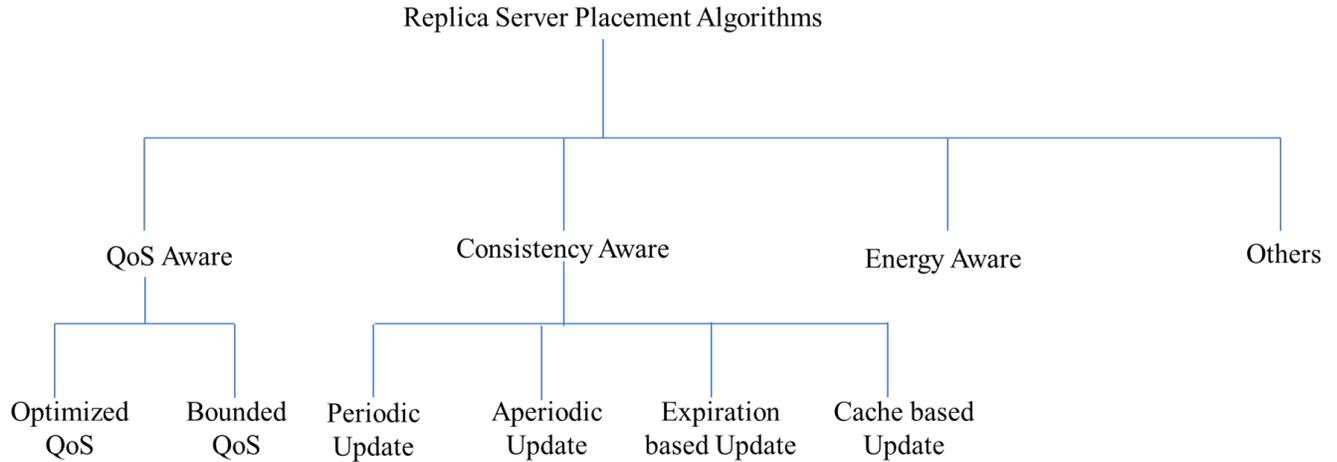

Fig. 6. Taxonomy of Replica Server Placement Algorithms for Traditional CDNs

This feature is in contrast to the considerations by existing algorithms on facility location problem and its generalizations, where the number of potential server locations ($M$) is much smaller than the number of end-user locations ($N$). Moreover, the GNP was used to compute latency with lower computational overhead. Another cluster based algorithm, called HotZone [25] is proposed to improve the computational cost and accuracy of replica server placement which considers the latency as the QoS metric. To avoid the high computational cost of greedy algorithms incurred as a result of all-pair latency estimation, the authors rely on GNP based latency estimation as described in Section III-B. Qiu *et al.* [61] propose various algorithms (random, Hotspot and Greedy) to solve the *uncapacitated K-median* model of replica server placement. The QoS metric to be optimized includes hop-count or latency. The random algorithm selects $K$ replica servers at random from a set of potential locations. This is the simplest form of replica server placement algorithm and can be used when there is no information on the inputs (e.g. topology, cost, latency, end-user load, etc.). The Hotspot algorithm is yet another primitive algorithm that selects $K$ replica servers based on only the end-user load information. The Greedy heuristic works in an iterative way by selecting a replica server that adds the lowest incremental cost to the set of already selected replica servers. The greedy heuristic terminates when $K$ number of replica servers are selected. Another greedy algorithm is proposed [47] to place replica servers in an adaptive multimedia system (AMS) with optimized QoS. The system dynamically places replica servers near the end-user locations in order to improve the QoS (i.e. latency) of video delivery. The greedy algorithm is similar to the greedy algorithm proposed by Qiu *et al.* [61]. However, unlike Qiu *et al.* [61] , it minimizes the sum of latency in updating the replica servers in addition to the latency in delivering content to the end-users. Radoslavov *et al.* [53] also proposes two heuristics based on node degree to place replica servers with the objective of minimizing latency.

Replica server placement problem with optimized QoS has also been addressed in Wireless CDN. In a wireless CDN, CDN entities, such as, the origin server, the replica server and the end-users, communicate with each other through wireless links. Typical examples include content delivery in mobile networks and ad-hoc wireless networks. Sung *et al.* [46] address replica server placement in wireless CDN. The replica server placement problem is modelled as *uncapacitated facility location* model and ILP is presented. Several QoS metric, such as, goodput, hop-count, interference and latency are combined to form the objective function. No algorithm is however discussed to solve the model.

*2) Bounded QoS:* Nguyen *et al.* [54] propose heuristics based on Lagrangian relaxation and greedy approach to solve the optimal resource provisioning problem. The problem is a joint optimization problem of replica server placement and content placement. The problem is formulated as Mixed Integer Linear Programming (MILP). An important and practical assumption made by the authors in [54] is that the requests generated from each end-user to access a content object are split among different replica servers having a copy of the object. Hence the replica server placement problem is an instance of the *splittable facility location* model. The cost to be optimized includes deployment cost and delivery cost. The QoS bound is represented using the maximum value of average distance between the end-user and a replica server. The authors [54] use a greedy heuristics based on two-level search process to find the placement of replica servers and the best replication of content objects.

A dynamic placement of replica servers is proposed by Chen *et al.* [33], where the problem involves finding minimum number of replica servers while satisfying QoS and servers' capacity constraints. A smart algorithm is proposed to solve the problem. Hei *et al.* [45] propose an optimal (called SR-optimal) as well as a greedy algorithm (called SR-greedy) to address the placement of streaming servers with bounded QoS. Bandwidth is considered as the QoS parameter. The placement problem is formulated using MILP. The problem involves the determination of the number of servers, location of servers and amount of Internet access bandwidth, which is bounded by the maximum throughput of the network interface. The objective function in the MILP formulation includes the deployment cost, the delivery cost and the network traffic. A heuristic called Server placement (SP) is proposed [48] to solve the streaming server placement problem in media streaming CDNs that use multiple descriptions coding to achieve reliable streaming. Using path diversity (i.e. by transmitting frames over different paths) packet loss rate for the end-user is reduced. Two QoS metrics: latency and reliability are considered. This streaming server placement problem is formulated as MILP.

Rodolakis *et al.* [27] present pseudo-polynomial and polynomial algorithms to solve the replica server placement problem. The problem is modelled as *splittable soft capacitated facility location* model. The replica server placement model involves finding the optimal location of replica servers, the type of replica servers and the number of replica servers of each type and the assignment of end-users to a set of replica server in a way that the cost is minimized and QoS is satisfied. A combination of centralized greedy and distributed greedy algorithms are proposed in [66] to solve the replica server placement problem. The problem entails finding the minimum number of replica servers, such that each end-user can access one of the replica servers with a bounded QoS. The problem does not incorporate cost minimization as all replica servers have homogeneous deployment cost.

The most common assumption regarding end-user assignment in the replica server placement is that the end-user requests are served by their closest replica server. However, such an assumption has the drawback of inefficient resource utilization. The assumption is relaxed in the replica server placement models presented in [56], wherein the authors studied the complexity of replica server placement for hierarchical network topology with two different policies: Upward and Multiple. The upward policy allows the end-users to receive service from any replica server along the path towards the root in the hierarchy (i.e. a tree). In the multiple policy, the end-user workload is split among servers. Both policies allow for more effective utilization of resources for the replica servers. Their replica server placement is formulated using ILP and involves finding the number and location of the replica servers, such that the deployment cost is minimized and end-users are assigned to replica servers by satisfying the server capacity and QoS constraints. With homogeneous storage cost of replica servers and without QoS, the replica server placement problem with Upward policy turns out to be NP-hard; whereas the Multiple policy allows the problem to be solved in polynomial time. When QoS is enforced the placement heuristics corresponding to closest policy, upward and Multiple policies are called Closest Big Subtree First (CBS), Upward Minimal Distance (UMD) and Multiple Minimal Requests (MMR), respectively. To place various types of CDN servers such as cache servers, video compressors, video transcoders, etc., End Point Placement algorithm [67] is proposed to serve media sessions with lower cost and a QoS guarantee. A session is defined as a group of end-points. Examples of sessions include unicast communication, multicast group communication, etc. The CDN servers are to be placed on the nodes that belong to the route of the sessions.

The replica server placement problem is modelled as the variation of *number of centers* model i.e. to find minimum number of servers such that the QoS constraints of all sessions are satisfied. In the End Point placement algorithm, the first server is placed on the most frequently used node (i.e. the node through which highest number of sessions pass through) and place the first server on that node. Once a session is covered, it is eliminated from further consideration. This procedure continues until all sessions are covered. TABLE II shows summary of QoS aware replica server placement algorithms.

### B. Consistency Aware

Many replica server placement algorithms are proposed to ensure consistency of the replica servers. Primarily, there are four ways of achieving consistency: 1) Periodic Update: in which replica servers are periodically refreshed by the origin server or replica servers, 2) Aperiodic Update in which an update is disseminated when the content is updated in origin server (or replica server), 3) Expiration based update in which replica servers obtain an up-to-date copy from nearest replica server if their replica has expired and when a request is received from end-users, and 4) Cache based update in which the update is disseminated only when a cache miss occurs at replica servers.

*1) Periodic Update:* In [68], a consistency aware replica server placement algorithm is proposed for tree topologies. In the topology, each leaf node i.e. nodes where end-user requests are generated is associated with an access frequency that indicates the popularity of the replica stored at that node over a certain time period. The origin server periodically disseminates updates to all replica servers using a distribution tree to ensure consistency between the replicas and the origin copy. This cost to be minimized includes the delivery cost and update cost. The algorithm is designed using dynamic programming technique and called Distributed Popularity Based Replica Placement (DPBRP). The authors discuss the replica server placement problem with and without QoS constraints and also provide two versions of the DPBRP algorithm corresponding to with QoS and without QoS.

TABLE II
SUMMARY OF QoS AWARE REPLICA SERVER PLACEMENT ALGORITHMS

| Paper | Minimization Objective | QoS Optimized/ Bounded | Integer Prog. Formulation | Theoretical Placement Model | Algorithm Approach |
|---|---|---|---|---|---|
| Jamin *et al.* [24] | Latency | Optimized | NoF[6] | Minimum *K*-Center | *l*-greedy, Transit Node Heuristic |
| Jamin *et al.* [63] | Latency | Optimized | NoF | Minimum *K*-Center | *K*-HST based greedy placement |
| Bhulai *et al.* [43] | Hop Count | Optimized | MIP | NoP[7] | Greedy, Cluster-based Heuristic |
| Wu *et al.* [44] | Hop count | Optimized | NoF | NoP | Genetic Algorithm |
| Yin *et al.* [26] | Cost, Latency | Optimized | NoF | Uncapacitated Facility Location | Improved *K*-means clustering |
| Szymaniak *et al.* [25] | Latency | Optimized | NoF | Uncapacitated K-Median | HotZone (Cluster based) |
| Qiu *et al.* [61] | Latency/Hop Count | Optimized | ILP | Uncapacitated K-Median | Iterative Greedy, HotSpot, Random |
| Goldschmidt *et al.* [47] | Latency | Optimized | NoF | NoP | Greedy |
| Radoslavov *et al.* [53] | Latency | Optimized | NoF | NoP | Node degree based algorithm |
| Sung *et al.* [46] | Hop count, Interference | Optimized | ILP | Uncapacitated Facility Location | No Algorithm is discussed. |
| Nguyen *et al.* [54] | Cost | Bounded | MILP | Splittable Facility Location | Langrangian heuristic, Greedy heuristic |
| Hei *et al.* [45] | Cost, Network Traffic | Bounded | MILP | NoP | SR-Optimal, SR-Greedy |
| Ahuja *et al.* [48] | Number of Servers | Bounded | MILP | NoP | Server Placement (SP) Algorithm |
| Rodolakis *et al.* [27] | Cost | Bounded | NoF | Splittable Soft Capacitated Facility Location | Pseudo-polynomial and polynomial algorithm |
| Jeon *et al.* [66] | Number of servers | Bounded | NoF | NoP | Greedy |
| Benoit *et al.* [56] | Cost | Bounded | ILP | NoP | Closest Big Subtree First (CBS), Upward Minimal Distance (UMD) and Multiple Minimal Requests (MMR), |
| Choi *et al.* [67] | Number of Servers | Bounded | NoF | Number of Center | End point Placement |

[6] NoF: No formulation is provided.
[7] NoP: The theoretical models are either not applicable or no placement model is discussed.

In [69], a greedy based algorithm called greedy-cover is proposed. The replica server placement problem defined in [69] involves finding the number and location of replica servers so that delivery cost and update cost is minimized and QoS (i.e. latency) constraint is satisfied. The greedy-cover algorithm achieves improvements over *l*-Greedy-Insert [44], *l*-Greedy-Delete [44] and greedy-MDS [45] in terms of execution time and hence has more scalability over the related algorithms. Moreover, the performance of greedy-cover is significantly better than *l*-Greedy-Insert, *l*-Greedy-Delete and greedy-MSC with higher QoS restrictions (i.e. very stringent latency requirements). In the update method of [63], a multicast tree, also known as update distribution tree, is used to periodically disseminate updates from origin server to the replica servers. The multicast tree is constructed as shortest path tree that is rooted at the origin server and connects all replica servers to the origin server through shortest path links. During the update dissemination, each node in the update distribution tree receives the updates from its parent and distributes it further to its children. The dissemination continues until the updates are received by all replica servers.

Tang and Xu [70] propose two algorithms: Greedy Insert and Greedy Delete to address consistency aware replica server placement in general graphs and a dynamic programming based algorithm for hierarchical network topologies The replica server placement problem involves finding the number and location of replica servers, such that the deployment cost and update cost are minimized and any end-user can receive content within its QoS bound. The consistency is achieved by using multicast based update dissemination similar to the method used in [69].

Two service models: replica-aware model and replica-blind model are presented. In the replica-aware model, nodes are aware of the replica server location by maintaining the object identifiers. If the request cannot be served locally, the requested object is retrieved from a nearby replica server. In the replica-blind model, the replica server locations are not known to the nodes. As a result, the end-users are served only if the node receiving the requests holds a replica. The authors show that the replica server placement problem in the replica-aware service model is NP-complete for general graphs.

*2) Aperiodic Update:* In [55], a dynamic programming based algorithm was proposed to optimally place *K* replica servers in a hierarchical network with minimum deployment cost, delivery cost, and update cost. The replica servers are assumed to have bounded capacity. Update cost occurs when an end-user writes to an object, which results in a dissemination of the update to the rest of the replica servers holding a copy of the object. The update cost is obtained by the cost of a minimum spanning tree rooted at the object at which the first write has occurred. The authors define a residence set as the set of replica servers that contain a replica of the object. The optimal residence set problem is to find an optimal residence set that minimizes the delivery cost and update cost. Without update cost, the optimal residence set problem is similar to the capacitated facility location problem.

The optimal residence problem is proved to be NP-complete in a tree network by reducing it from the bin-packing problem. When only delivery cost is considered, the optimal residence set problem with the limit *K* on the number of replica servers is an instance of the *K-median* model.

In [42], optimal placement of *K* replica servers [42] in hierarchical network topologies is achieved using two algorithms: Aggregate Access (AGGA) and weighted Popularity (WPOP). Update operations are allowed on the content stored on the replica servers. The example applications include stock quote services and distributed ticketing. The replica server placement model presented in [42] involves minimization of the delivery cost and update (write) cost. Note that, the replica servers in this model are not full replicas rather partial replicas hosting a subset of the objects.

The AGGA algorithm is based on aggregate access rate (read rate and write rate) of each node in the network. First, an optimal residence set determination algorithm is used to find the optimal replica servers with a constraint *K* on the maximum number of replica servers to be placed. If *K* replica servers are found, the optimal placement is reached. On the other hand, if the number of replica servers returned is less than *K*, a greedy approach is used to select the rest of the replica servers by selecting one node at a time that adds the lowest incremental cost. On the other hand, the WPOP algorithm places replica servers on top nodes sorted according to their weighted popularity. The weighted popularity is calculated using total access rates (read rate and write rate) of each node and whether the node belongs to an optimal residence set [42]. In [71], another work on optimal placement of replica servers in a tree network is presented. The objectives include deployment cost and delivery cost.

*3) Expiration based Update*

There have been some algorithms ( [72], [73]) that address replica server placement under expiration based update. In this consistency scheme, each replica server is assigned a Time-To-Live (TTL) value. On expiration of the TTL value, the update request from a replica server is forwarded in the path towards the root until a replica server having a valid copy is reached. Tang and Chanson [73] designed a dynamic programming based algorithm to address the replica server placement. The cost to be minimized includes delivery cost and update cost. The update cost depends on the time to live value of the replica servers and the request arrival patterns.

*4) Cache based Update*

There are two main types of cache based update; cooperative and non-cooperative. In case of cooperative caching, if the content is not found (i.e. cache miss) in the replica server, it is retrieved from another replica server; whereas in non-cooperative caching, the content is retrieved from the origin server.

In [74] , a greedy heuristic was proposed to address the joint optimization of replica server placement, content replication and request routing. The authors consider cooperative caching scheme and compute the update cost accordingly. The joint optimization problem aims to minimize delivery cost and update cost. The authors show that the joint optimization problem is NP-hard by reducing it to uncapacitated facility location problem.

In [65], a dynamic programming based algorithm was proposed to address replica server placement in tree networks. The authors consider non-cooperative caching scheme and model the replica server problem as *K-cache location* model. The objective is to place *K* caches such that the delivery cost and update cost is minimized.

Yang *et al.* [75] investigated various algorithms to solve replica server placement problem considering that the non-cooperative caching scheme is used to update the replica servers. The objective of the model is to find optimal number and location of replica servers that minimizes the delivery cost and update cost. The capacity (in terms of number of objects) of replica servers is known and each replica server has certain number of requests from end-users. The authors use Zipf-like distribution to estimate the hit ratio of the obtaining content from replica servers. Without the capacity constraint, the replica server placement problem is similar to *K-cache* location model. Three existing algorithms such as greedy algorithms, max fan out algorithm and hot spot algorithm are modified to solve the model.

A greedy heuristic was proposed in [76] to solve replica server placement with non-cooperative caching scheme. The replica server placement problem is called as single server content distribution network problem (SCDNP), which involves finding the optimal number and location of replica servers, optimal placement of objects in the replica servers and assignment of end-users to appropriate replica server, such that, deployment cost, delivery cost and update cost are minimized. The update cost is incurred when the replica server does not hold the requested object and hence receives it from the origin server. The SCDNP is modelled as Integer Non-Linear Programming (INLP) problem. This problem is NP-hard as it is an instance of *multicommodity uncapacitated facility location*[8] problem. Bender's decomposition approach is used to find the optimal solution for small-scale CDNs; whereas the greedy-heuristic is used to solve SCDNP in large-scale CDNs.

TABLE III shows summary of consistency aware replica server placement algorithms.

### C. Energy Aware

In CDNs, power consumption of the replica servers is one of the important factors that affect the placement and replacement of replica servers. In other words, the replica servers must be placed in a way that minimizes the energy cost. An energy aware replica server placement algorithm was proposed in [77], which uses dynamic programming technique to find the optimal number and location of replica servers in a tree network to minimize energy cost. The end-users are considered as the leaf nodes and are served by internal nodes (candidate for replica servers) in the tree. For flexible power management, a multi-modal processor model is considered, which allows replica servers to switch between multiple processor modes with different speeds, thereby the power usage varies drastically among replica servers.

When there is no replica server in the network, the replica server placement problem involves finding new replica servers to be added to the network. However, when the network already contains some replica servers, the replica server placement problem involves finding new replica servers to be added, existing replica servers to be deleted and existing replica servers to switch processor modes. The authors model the replica server placement problem using a bi-criteria formulation, Minpower-Bcost, which is to minimize the energy cost while satisfying a bound on the cost.

The cost function is a combination of static cost incurred in turning on/off the server and a dynamic cost (addition, deletion, switching processor modes). Two versions of this problem are presented, Minpower-Bcost-WithPre and Minpower-Bcost-NoPre, corresponding to whether the network contains pre-existing replica servers or not, respectively. Several other independent problems, such as, Minpower (minimizing power consumption cost), and Mincost (minimizing configuration cost) are presented. In Mincost problem, one mode is considered and the cost involves the configuration cost in adding and/or deleting servers.

In [78], an extension of the above model is discussed. The model involves determining location of replica servers and distributing end-users requests among the replica servers so that the total power consumption is minimized. The Mincost problem [77] is also addressed in [79], wherein the authors propose greedy heuristics and Tabu search based heuristics to find near-optimal solution.

Energy aware placement of replica servers have also been addressed in CDNs deployed over telecom infrastructures. In Mandal *et al.* [34], the replica server placement is analyzed for both centralized (e.g. data centers) as well as distributed scenarios (e.g. cache nodes in the core network). These scenarios result in different amount of energy consumption in storing and transmitting content. Data centers incur higher transmission energy as they are far from end-users, but incur lower storage energy. On the other hand, cache nodes are closer to end-users and hence reduce the transmission energy consumption. However, as the cache nodes are deployed in large numbers, they incur higher energy consumption due to storage. In [34], the authors defined a placement problem that involves minimizing the total energy consumption by determining an appropriate placement of replica servers and assigning end-users to replica servers. The problem is formulated as ILP and an algorithm called Popularity Aware Content Replication (PACR) is proposed that considers the popularity of content objects to solve the problem. Two other algorithms, called On-way Replication and Weighted Cost Replication are presented to reduce energy by dynamically adjusting the placement of content according to the temporal variation in end-user activity.

In [80], a joint content source selection (i.e. assignment of end-users to the content location) and download route selection (i.e. selection of a routing path to download content) problem is addressed to minimize energy consumption in hybrid CDN–P2P systems. In a P2P system, end-users cache contents to serve other end-users.

---

[8] Multicommodity facility location model is an extension of facility location model where end-users have demand for different objects.

However, some P2P-based approach such as nano-data-center-based approach has been studied to consume more system-wide energy consumption than CDNs. As a result, energy-efficiency is a key issue in hybrid CDN-P2P system. The joint content source selection and download route selection problem has two sub-problems. The first sub-problem decides whether to select replica server or peers as the content source for a given end-user request. The second sub-problem involves finding an energy-efficient routing path for downloading the content. The problem is formulated as ILP. Two algorithms called, Minimize Server Bandwidth (MSB) and Minimize Instantaneous Energy (MIE). The MSB algorithm selects a peer as the content source when sufficient peers are available in order to reduce the bandwidth usage of replica servers. In MIE algorithm, the peer that consumes least amount of energy is selected.

TABLE IV shows summary of the literature on energy aware replica server placement algorithms.

TABLE III
SUMMARY OF CONSISTENCY AWARE REPLICA SERVER PLACEMENT ALGORITHMS

| Paper | Consistency Mechanism | Minimization Objective | Integer Prog. Formulation | Theoretical Placement Model | Algorithm Approach |
|---|---|---|---|---|---|
| Shorfuzzaman *et al.* [68] | Periodic | Cost | NoF | NoP | Dynamic Programming |
| Wang *et al.* [69] | Periodic | Cost | NoF | NoP | Greedy-Cover |
| Tang and Xu [70] | Periodic | Cost | NoF | NoP | *l*-Greedy |
| Kalpakis *et al.* [55] | Aperiodic | Cost | NoF | Uncapacitated K-Median | Dynamic Programming |
| Xu *et al.* [42] | Aperiodic | Cost | NoF | NoP | Aggregate Access (AGGA) and weighted Popularity (WPOP) |
| Tang and Chanson [73] | Expiration based | Cost | NoF | NoP | Dynamic Programming |
| Lim *et al.* [74] | Cache based (Non-Cooperative) | Cost | INLP | Uncapacitated Facility Location | Greedy Heuristic |
| Yang *et al.* [75] | Cache based (Non-Cooperative) | Cost | NoF | *K*-Cache Location | Modified Algorithms (Greedy, node degree based, Hot Spot) |
| Krishnan *et al.* [65] | Cache based (Non-Cooperative) | Cost | NoF | *K*-Cache Location | Dynamic Programming |
| Bektas *et al.* [76] | Cache based (Non-Cooperative) | Cost | INLP | Mulicommodity Uncapacitated Facility Location (multiple objects)/Uncapacitated Facility Location (Single object) | Greedy Heuristic |

TABLE IV
SUMMARY OF ENERGY AWARE REPLICA SERVER PLACEMENT ALGORITHMS

| Paper | Minimization Objective | Integer Programming Formulation | Theoretical Placement Model | Algorithm Approach |
|---|---|---|---|---|
| Benoit *et al.* [77] | Cost | NoF | NoP | Dynamic Programming |
| Aupya *et al.* [78] | Cost | MILP | NoP | Greedy Heuristic, Speed Heuristic, Excess Heuristic |
| Wang *et al.* [79] | Cost | NoF | NoP | Greedy Heuristic, Tabu Search |
| Mandal *et al.* [34] | Energy Consumption | ILP | NoP | Popularity aware Content Replication(PACR), On-way Replication, Weighted Cost Replication |
| Mandal *et al.* [80] | Energy Consumption | ILP | NoP | Minimize Server Bandwidth, Minimize Instantaneous Energy |

TABLE V

SUMMARY OF OTHER REPLICA SERVER PLACEMENT ALGORITHMS

| Paper | Minimization Objective | Integer Programming Formulation | Theoretical Placement Model | Algorithm Approach |
|---|---|---|---|---|
| Bassali *et al.* [81] | Cost | NoF | Uncapacitated K-Median | Node degree and hop count based algorithms |
| Varadarajan *et al.* [82] | Cost | MINLP | Uncapacitated *K*-Median | Greedy |
| Thouin *et al.* [83] | Cost | MINLP | NoP | 2-step Search Heuristic |
| Laoutaris *et al.* [58] | Cost | NoF | Uncapacitated *K*-Median, Uncapacitated Facility Location | Iterative Local Search |
| Luss *et al.* [84] | Cost | ILP | NoP | Multistate Dynamic programming (DP) |
| Laoutaris *et al.* [85] | Cost | ILP | Multicommodity K-Median | Greedy Heuristic |
| Ho *et al.* [86] | Cost | ILP | Uncapacitated K-Median | Optimal Algorithm |

*D. Others*

In [81], four algorithms were proposed for placing replica servers in suitable ASs in an AS level topology. The objective is to place replica servers so that the average delivery cost is minimized. The replica server placement model is close to the *uncapacitated K-median* model. The first algorithm is called highest-degree-first and it places replica servers in ASs having higher node degrees. This is a reasonable metric as it results in higher possibility of reaching more end-users and hence minimizes the average latency. [53]. The second algorithm is called farther-first and it places *K* replica servers in ASs in a way that all replica servers have higher degrees and no two replica servers are in close proximity of each other. The third algorithm is called hybrid algorithm which accepts a parameter *n* (< *K*) as input. It executes highest-degree-first for placing n replica servers and then executes farther-first to place the remaining *K-n* replica servers. The fourth algorithm is called optimized-hybrid which runs hybrid algorithm for *n*=1, 2,…*K* and selects the placement that minimizes the average delivery cost.

Greedy based heuristics were proposed in [82] to solve the replica server placement problem. The replica server placement problem is modelled as a variant of *uncapacitated K-median* model. The objective function includes delivery cost. The problem is formulated as Mixed Integer Non-Linear Programming (MINLP) problem. The novelty of this work is a host coverage metric (see paper [82] for details), which provides a long-term measure of the end-user workload and hence negates the impact of fluctuating workload on the cost and performance due to a given placement. In the first heuristic i.e. greedy-exchangeCompute, host coverage and distance are considered simultaneously; whereas, the second heuristic i.e. greedy-coverageCompute considers both metrics independently.

In [83], a 2-step search heuristic is designed to find the near-optimal placement of VoD servers, a problem referred to as VoD Equipment Allocation problem (EAP). The network is represented as a logical star topology, where the nodes include end-users and the origin server, which is at the center of the topology. All nodes represent the potential locations for placing the VoD servers. Each end-user node is a cluster of real end-users and the clusters are obtained by dividing a Metropolitan Area Network (MAN) into several zones. The VoD servers are of various types and act as both streaming and replica servers. The placement involves finding the optimal number and type of VoD servers and the fraction of load allocated to each of them. A multi-model variant of the EAP allows placement of multiple VoD servers of different types at each location. The objective function includes deployment cost and delivery cost. The EAP is modelled as a MINLP problem.

Laoutaris *et al.* [58] present a distributed algorithm for finding placement of replica servers with minimized cost. The motivation to design distributed approach came from the fact that real-time collection of global topology information and the end-user workload information for a large-scale network creates huge overhead unless the CDN provider relies on third party to obtain this information. The distributed algorithm is designed based on iterative local search method. It is scalable and allows addition/deletion of replica servers without incurring too much overhead in exchanging the topology and end-user demand information. The replica server placement is mapped to the uncapacitated *K*-median (UKM) and the uncapacitated facility location (UFL) and the exact problem to solve is selected based on the specific information needed i.e. number of replica servers in UKM and the deployment cost in UFL.

In [84], the authors proposed multistate dynamic programming based algorithm to solve a joint optimization problem. The problem involves deciding the optimal placement of the replica servers, optimal placement of objects in the replica servers and assignment of the end-users to appropriate replica server in a tree network. The objective is to minimize the deployment cost (installation cost, storage cost and processing cost) and delivery cost. The joint optimization problem is modelled as an Integer Linear Programming problem.

In Laoutaris *et al.* [85], a greedy heuristic is proposed to find solution to a joint optimization of replica server placement and object placement for tree networks. The placement model involves minimization of delivery cost and satisfaction of replica server capacity. It is formulated using ILP and is close to *multi-commodity K-median* model[9].

Ho *et al.* [86] present a replica server placement model that is robust to variations in traffic volumes. Uncertainty in traffic demand has a significant impact on CDN performance as different scenarios result in different traffic volumes in CDNs. As a result, with a given placement of replica servers, CDN providers may have to deal with high cost in delivering content to end-users in worst-case traffic scenarios, such as flash crowds. Ho *et al.* [86] present a robust replica server placement model that minimizes the delivery cost across a variety of traffic demand scenarios and minimize the deviations from the optimal solution for each scenario. For a single scenario, the model is reduced to *uncapacitated K-median* model. To find a placement, two criteria are employed: minimax and minimization of relative regret. The first criterion aims at achieving the best out of all worst possible scenarios. The motivation for this criterion come from the fact that the CDN provider may not want good solution for a specific scenario, but rather wants a solution that performs reasonably well across a multitude of scenarios. The second criterion aims at finding a solution close to the optimal solution. These two criteria are conflicting resulting in the pareto-optimal solutions. The authors presented the robust replica server placement using ILP where the second criterion is modelled as a constraint. Optimal algorithm such as branch and bound is used to solve the ILP. The simulation study [86] shows that the scenario-based replica server placement outperforms the replica server placement approach with deterministic traffic demand. All replica server placement algorithms in the "Others" category are summarized in TABLE V.

## VI.  REPLICA SERVER PLACEMENT ALGORITHMS FOR EMERGING PARADIGMS BASED CDN

In this section, we discuss various replica server placement algorithms proposed for emerging paradigms based CDNs. First we discuss algorithms for cloud based CDNs followed by algorithms for NFV based CDN.

### A.  Cloud based CDN

The cloud based CDN algorithms are classified into four main categories: 1) QoS Aware, 2) Consistency Aware, 3) Others.

1)  *QoS Aware*: In [87], optimal allocation of cloud resources is addressed for cloud based video services. In particular, they determine dynamically the minimum number of servers in a data center to meet the latency requirements of VoD and Live streaming services. The authors use a prediction method to anticipate the load in the near future and instantiate Virtual Machine (VMs) to meet the service requirements. Clearly, live streaming has stringent latency requirements than VoD. The VoD servers are allowed to serve the sessions at a faster rate before a burst of live streaming request is detected, after which the VoD servers are delayed in order to allow the live streaming servers to fully utilize the bandwidth resources.

Zhang *et al.* [88] propose an online algorithm based on Model predictive control approach to solve the replica server placement problem by considering dynamic pricing of resources and fluctuating end-user workload. Dynamic pricing is often employed as a result of varying electricity costs in regional power systems. The cloud providers vary the resource prices to obtain stability in their profit margins. The replica servers are deployed on data centers of multiple cloud providers. A discrete-time system model is considered where time is divided into a number of intervals called reconfiguration periods. The duration of a reconfiguration period indicates the periodicity with which the replica server placement is performed. The authors in [88] aim to find the optimal number of servers in data centers and optimal assignment of end-users to the servers during a time period so that the operational cost is minimized and QoS (i.e. latency) is satisfied. The optimal number of servers is determined by applying optimal reconfiguration (i.e. number of servers to be added/deleted) to the existing configuration (i.e. number of existing servers). The operational cost includes the reconfiguration cost in addition to the leasing cost. The reconfiguration cost includes the switching cost of set up (i.e. distributing VM images when new replica servers are added) and the tear down cost (i.e. the cost of transferring data/state of VMs when replica servers are deleted). The placement is formulated as a MINLP problem.

In [89], an optimal algorithm is proposed to address optimal resource provisioning for cloud based video streaming system. The objective is to find optimal amount of leased resources (i.e. VM instances) to meet QoS in the wake of fluctuating end-user workload, while minimizing the operational cost. The operational cost includes the cost of leasing VM instances to be used to deploy the streaming servers. The authors [89] adopt three pricing models of Amazon (on-demand, reservation and spot) to model the cost of resource provisioning. The end-user demand for certain video content/channel is predicted by considering the video popularity of the content/channel. They introduce a QoS metric, overload probability, which is the probability that the total resource (bandwidth) demand exceeds the capacity of the leased resources. The optimal resource provisioning problem is modelled using ILP.

Ferdousi *et al.* [90] present a data center placement and content placement that ensures content availability in case of disasters or targeted attacks on data centers. ILP is used to model a static data center placement and content placement, where the objective is to minimize risk defined as the expected content loss is minimized. The authors consider one of the two cases that result in loss of access to content due to disaster in determining expected content loss. In the first case, the content cannot be accessed if the data center is not available. In the second case, the content cannot be accessed if the data center is not reachable due to failure of the nodes and links on the network that connects the end-user to the data center. In the latter case, the QoS requirements of end-users are used to determine expected content loss. The solution to ILP is used as input to a dynamic heuristic, called Disaster-Aware Dynamic Content Management (DADCM) that reconfigures the initial placement of contents in order to reduce risk under evolving network and disaster scenarios.

---

[9] Multicommodity version of K-median model refers to an extension of K-median model where end-users have demand for different objects.

*2) Consistency Aware:* Rappaport and Raz [28] model the replica server placement problem in cloud based CDN as soft capacitated connected facility location model and propose an approximation algorithm to solve the problem. The model aims to minimize the operational cost that includes the leasing cost, the delivery cost and the update cost. The solution to the model contains replica server locations and an optimal tree that interconnects the replica servers to ensure consistency. Moreover, since the placement minimizes the update traffic between replica servers, it results in reduction of the inter-data center traffic, which in turn reduces the load on the network resources between data centers. The authors in [28] also define a $\lambda$-loaded facility location problem by introducing a constraint that the load served by each replica server must be above a predefined threshold. The approximation algorithm for the soft capacitated facility location problem is designed using approximation algorithm for the uncapacitated facility location problem and approximation algorithm for minimum Steiner tree problem. The proposed algorithm is offline in nature.

In [91], the optimal CDN provisioning problem is presented to find the optimal configuration for video CDN. The video CDN uses layered cache servers. The optimal configuration involves the number of cache servers, size of the caches and the amount of peering bandwidth needed to serve the end-users. The objective is to provision cache servers in a way that minimizes the operational cost (i.e. storage and bandwidth costs). The CDN provisioning problem is formulated as an INLP problem and an optimal algorithm is designed to find the optimal configuration.

*3) Others:* Wang *et al.* [92] propose an enhanced Depth First Search (DFS) algorithm to find optimal lease schedule in order to accommodate spatial and temporal variations in end-user demands. The optimal lease schedule shows the appropriate server to lease at a given time instant to serve a predicted user demand. The authors incorporate locality awareness, that is, regional distribution of end-users into the placement decisions. The latency is reduced by leasing local cloud servers to serve as many end-users as possible. This also reduces the cross-region traffic. The algorithm minimizes operational cost (i.e. leasing cost) and cross-region traffic subject to QoS constraint. The two objectives: minimization of operational cost and minimization of cross-region traffic are conflicting with each other as the more the number of servers leased locally to serve maximum number of end-users, the higher the leasing cost. The cost function is thus obtained by a linear combination of objectives expressed as a ratio between

[0, 1]. The DFS algorithm is offline in nature. To improve the search efficiency and reduce the computation time of the proposed algorithm, they sort the cloud servers in the ascending order of the leasing cost, thereby allowing the algorithm to obtain near-optimal solutions in less time.

Chen *et al.* [29] present multitude of greedy algorithms to solve replica server placement problem in cloud based CDN. Their objective is to minimize cost, while satisfying QoS (i.e. latency) of end-users and connecting every end-user to at least one replica server. The cost includes the server deployment cost (i.e. storage cost) and update cost (updated from origin server only). The network cost of transferring content from source to destination involves the cost of downloading and uploading content from and to the cloud. This is in contrast to traditional CDN, where the network cost involves a single cost in transferring content from source (e.g. origin server) to a destination (e.g. replica server). Chen *et al.* [29] provide an ILP for replica server placement problem which is an instance of uncapacitated connected facility location. They present both offline (Greedy Site and Greedy User) and online algorithms (e.g. Greedy Request Only and Greedy Request with Pre-allocation). The Greedy Site is based on set covering algorithm. On the other hand, in Greedy User, for each end-user the replica server that incurs the lowest cost and satisfies QoS is placed. Note that offline algorithms pre-deploy replica servers before the end-user request is received. The Greedy Request Only online algorithm executes Greedy User to deploy a suitable replica server based on the first request. However, the time needed to deploy may violate QoS. Hence, once the first request is received, if the chosen replica server is not yet placed, the request is served from the origin server. Greedy Request Pre-allocation reduces the number of QoS violations by pre-deploying some servers based on knowledge of recent end-user requests.

Another greedy based replica server placement algorithm called Social Network Analysis (SNA) Inspired Greedy Virtual Surrogate Placement (SNA-GVSP) is proposed by Papagianni *et al.* [30]. It is assumed that CDN provider plays the role of cloud provider. The replica server placement problem is to find number and location of replica servers, such that operational cost is minimized and QoS requirements of end-users are satisfied. The operational cost involves the server deployment cost (i.e. storage cost), delivery cost and update cost. The unit storage cost is defined as a piecewise linear and convex function of the utilization of the disk resources over a time window at the cloud site. The replica server placement is formulated using MILP. The problem is similar to capacitated connected facility location model.

The SNA-GVSP algorithm is based on a centrality metric motivated by SNA concept. In particular, SNA-GVSP first prioritizes the cloud sites based on the centrality metric i.e. Shortest Path Betweeness Centrality $v$(SPBC) metric, which is the ratio of the number of shortest paths traversed through a node to the total number of shortest paths in a network. The SNA-GVSP algorithm then assigns end-users to a cloud site that maximizes the average SPBC of the set of already selected sites. The algorithm is executed in an offline manner.

Optimal deployment of video distribution services is addressed in [93]. Video services are bandwidth-intensive services. The authors in [93] assume that the video service provider (i.e. CCDN provider) leases the resources from multiple cloud providers in order to deploy the service in geographically dispersed data centers and improve the end-user QoS. The objective function includes the operational cost and latency. Since the costs of using other resources (e.g. CPU, memory) are negligible in comparison to bandwidth cost, only the latter represent the operational cost. The bandwidth cost is in the form of a non-decreasing concave function meaning the more you buy bandwidth units, the cheaper is the unit price, whereas latency is defined as a convex function. The cost and latency minimization is tackled by an offline algorithm designed based on Nash bargaining solution. The offline algorithm considers the predicted end-user workload in the future time slots. An online algorithm is also designed to minimize cost and probability of under-provisioning of resources. The online algorithm considers the predicted end-user workload at the beginning of a time slot and performs the adjustment or redeployment of resources in the data centers.

An adaptive replica server placement is presented in [94]. The objective of the placement is to find the optimal number and the location of servers, such that the operational cost and the number of reconfigurations are minimized. The cost involves deployment cost and delivery cost. The number of reconfigurations is bounded by incorporating a policy, which allows a certain number of reconfigurations (i.e. change in number of replica servers) for a given decrease in total operational cost from the previous placement. Their replica server placement problem is formulated as an integer linear programming problem. Their replica server placement problem is a variation of uncapacitated facility location problem. The authors consider that replica server placement must be conducted at regular intervals to adapt to the dynamicity in end-user workload. The motivation behind minimizing the number of reconfigurations is to avoid the cost and the time in placing or removing the replica servers. Note that, the reconfiguration might disrupt the service unless the incoming end-user requests are redirected to some existing servers. To quantify reconfiguration at a time instant $t$, the authors use hamming distance between the optimal set of replica servers at $t$-1 and every potential set of servers. It is shown that any approximation algorithm for facility location problem can be used to find the solution. In the simulations, the authors used a greedy based approximation algorithm with approximation ratio 1. 861 [95] to find the solution of their replica server placement problem.

TABLE VI presents the summary of works on replica server placement problem for Cloud based CDN.

TABLE VI

SUMMARY OF REPLICA SERVER PLACEMENT ALGORITHMS FOR CLOUD BASED CDN

| Paper | Objective | Integer Prog. Formulation | Theoretical Placement Model | Algorithm Type | Algorithm Approach |
|---|---|---|---|---|---|
| Rappaport and Raj [28] | Op. Cost | NoF | Soft Capacitated Connected Facility Location | Offline | Constant Factor Approximation |
| Wang *et al.* [92] | Op. Cost, Network traffic | NoF | NoP | Offline | Enhanced DFS |
| Chen *et al.* [29] | Op. Cost | ILP | Uncapacitated-Connected Facility Location | Offline, Online | Greedy |
| Papagianni *et al.* [30] | Op. Cost | MILP | Capacitated-Connected Facility Location | Offline | Greedy |
| Aggarwal *et al.* [87] | No. of servers | NoF | NoP | Online | Optimal Algorithm |
| He *et al.* [93] | Op. Cost, Latency | INLP | NoP | (Offline, Online) | Nash bargaining solution |
| Zhang *et al.* [88] | Op. Cost | INLP | NoP | Online | Model Predictive Control |
| Zhenghuan *et al.* [89] | Op. Cost | ILP | NoP | Online | Optimal Algorithm |
| Tran *et al.* [94] | Op. Cost, No. of Reconfigurations | ILP | Uncapacitated Facility Location | Online | Greedy |
| Mokhtarian *et al.* [91] | Op. Cost | INLP | NoP | Offline | Optimal Algorithm |
| Ferdousi *et al.* [90] | Risk | ILP | NoP | Offline | Disaster-Aware Dynamic Content Management (DADCM) Heuristic |

TABLE VII
SUMMARY OF REPLICA SERVER PLACEMENT ALGORITHMS FOR NFV BASED CDN

| Paper | Objective | Integer Programming Formulation | Theoretical Placement Model | VNF Type | Algorithm Type | Approach |
|---|---|---|---|---|---|---|
| Cohen *et al.* [31] | Op. Cost | ILP | Capacitated/Uncapacitated Facility Location | Not specific to CDN | Offline | Approximation Algorithm of Generalized Assignment Problem |
| Llorca *et al.* [32] | Op. Cost | MILP | Min cost Mixed-cast Flow (See [30]) | $v$Cache | Offline | Optimal Algorithm (branch and cut) |
| Bouten *et al.* [41] | Op. Cost | INLP | NoP | $v$Transcoder, vCache | Offline | Genetic Algorithm |
| Mangili *et al.* [96] | Op. Cost | SMIP | NoP | $v$CDN replica server | Offline | Greedy |

## B. NFV based CDN

In NFV based CDN, various algorithms have been proposed to place VNFs on NFVI nodes (e.g. network nodes or data centers) and assignment of end-users to the VNFs. In addition to VNF placement algorithms, there is one algorithm that addresses a joint placement of physical servers as well as VNFs in the CDN infrastructure. Most of the algorithms [32] [41] [96] [97] in proposed for NFV based CDN aim to minimize the operational cost. In addition to cost minimization, some VNF placement algorithms consider the QoS requirement (e.g. the tolerable latency in receiving the content) of end-users when placing the VNFs. Moreover, some other algorithm [98] attempts to optimize the network resources (e.g. bandwidth) when the traffic flows have different service requirements that enforce same set of VNFs to be executed in different order.

Llorca *et al.* [32] propose a joint optimization of content placement and $v$Cache server placement in SD$v$CDN (Software-Defined $v$CDN), which is a virtual cache network deployed over a distributed cloud network infrastructure. Content placement involves deciding the distribution of contents among $v$Cache servers and the $v$Cache server placement involves deciding the optimal number of $v$Cache servers at each cloud location. Each cloud location has capacity to host a given number of $v$Cache servers. The problem involves deciding the optimal content placement and optimal $v$Cache server placement, such that the total operational cost is minimized and the QoS for each object is satisfied. The operational cost is the sum of deployment cost (i.e. storage cost) and the delivery cost. Each cost comprises of a fixed operational cost associated with activation of resources without any load and a variable operational cost that depends on the usage of the resources based on the load. The problem is modelled as a minimum cost mixed-cast (unicast and multicast) flow problem with resource (storage and network link) activation decisions using MILP. Existing optimal algorithm (branch and cut) algorithm is shown to solve the problem.

The authors in [41] address a placement problem which involves optimally deploying data centers in an ISP network and deciding the number of VNFs (e.g. virtual Transcoder ($v$Transcoders) and virtual CDN Cache nodes ($v$CDN Cache)) in the data center. $v$Transcoders and $v$CDN Caches can be placed in an ISP's network close to the end-users to reduce the backhaul traffic. The Point-of-Presences (PoPs) of ISPs are distributed using a multi-layer network architecture, which comprises of four PoP layers: inner core, outer core, aggregation and access layer. As discussed in [41], there can be two strategies for deploying data centers. One approach is to deploy a few yet large sized data centers in the core. They can serve large number of users, but induces load on the backhaul network. Another approach is to place multiple small data centers in the access layer. This approach reduces the end-user latency along with network load. Note that, the second approach results in higher capital and operational costs. For example, since the datacenter locations are geographically dispersed, the costs for security, maintenance, etc., increases, since multiple smaller sites need to be maintained. Thus, the data centers must be placed in a way that simultaneously minimizes the cost and network load. The data center placement problem is formulated using INLP. The objective function includes a capital cost (i.e. cost of deploying data centers) and operational cost (i.e. delivery cost and update cost). Latency and link capacity are considered as constraints. Since, $v$CDN cache nodes are responsible for caching content, cache hit ratio is considered to compute the update cost. The authors [41] assume that cache hit ratio in a $v$CDN cache is proportional to its storage capacity. A Genetic Algorithm is used to find the optimal placement.

Authors in [96] address the placement of physical replica servers[10], where the CDN infrastructure also comprises of virtual replica servers[11]. It is considered that a fixed number of virtual replica servers are deployed to address the uncertain traffic. The traffic demand is modelled using stochastic process, where each traffic scenario occurs with a certain probability. The optimal placement of physical CDN replica servers is achieved by minimizing the overall cost and satisfying constraints such as capacity of the replica servers and maximum latency for a certain number of flows. The cost of placing physical CDN replica servers includes both capital cost (i.e. deployment cost) and operational costs, whereas, the cost of virtual CDN node includes only the operational cost. The optimal placement problem is formulated using Stochastic Mixed Integer Programming (SMIP). A greedy algorithm is designed to find optimal placement of physical and virtual replica servers. TABLE VII present the summary of the replica server placement algorithms for NFV based CDN. Note that all algorithms in NFV based CDN are offline in nature. However, they can be modified to act as online algorithms.

---

[10] No virtualization technology is used.
[11] Replica servers deployed as VNFs using NFV paradigm.

The authors in [97] introduce NFV location problem and propose approximation algorithm. The NFV location problem refers to placing one or more VNFs that form a service delivered to end-users. In some cases, the service is distributed among nodes, meaning VNFs are placed on VMs located at more than one node and the routing between VNFs is managed by SDN mechanisms. The NFV location problem involves finding the optimal number and the location of VNFs in the network and assignment of end-users to their required VNFs, such that the operational cost is minimized. The operational cost includes deployment cost, that is, cost of installing VNFs on the VMs and delivery cost. The NFV location problem is presented using ILP. It shares characteristics with facility location problem as well as Generalized Assignment problem (GAP). Both capacitated and uncapacitated versions of the NFV location problem are presented. The capacitated and uncapacitated version denote the fact that each VNF instance can serve up to a limited and unlimited number of end-users, respectively. The approximation algorithm is designed based on the on the approximation algorithms for GAP.

The VNF placement problem presented in [98] involves finding proper location of VNFs (e.g. Firewall and Video Optimizer) that minimize the total network bandwidth consumption, satisfies the service requirement of all traffic flows. The candidate locations include data centers and NFV capable nodes such as routers and switches. Service requirement of a traffic flow is expressed as set of VNFs and a chain that indicates the sequence in which the flow is processed by VNFs along the chain. The VNF placement problem is formulated as ILP.

## VII.  COMPARISON OF REPLICA SERVER PLACEMENT ALGORITHMS

In this section, we identify the requirements for an efficient replica server placement heuristic and then provide comparison of existing heuristics.

### A.  Requirements

*1) Cost Minimization:* This is the first and foremost requirement that must be incorporated by a replica server placement algorithm. In CDNs, the most important cost components are deployment cost, delivery cost and update cost. The deployment cost is a one time cost in traditional CDN unless the heuristic supports addition of new replica servers. On the other hand in Cloud based CDN, the deployment cost is basically the cost of leasing virtual resources from cloud providers. For private cloud, the deployment cost is replaced by a recurring maintenance cost. For VNF placement in NFV based CDN, the deployment cost may indicate the storage cost and/or the cost of installing VNFs; whereas data center placement and physical replica server placement involves a one-time deployment cost. Unlike deployment cost, the delivery cost and update cost remain operational cost incurred by using network resources for traditional CDNs as well as CDNs based on emerging paradigm. Update cost is sometimes not applicable to cloud based CDN and NFV based CDN when the servers to be placed do not store replica and rather possess different functionalities (e.g. media processing). Since, the above cost components reflect the primary activities in CDN, they must be considered by a replica server placement heuristic in order to provide CDN providers with significant cost savings.

*2) Bounded QoS:* End-user satisfaction in terms of providing a guaranteed QoS is also one of the primary goals of a CDN provider. The algorithm needs to consider QoS threshold for end-users that will receive service from one or more replica servers. It is important to note that, end-users on noticing service degradations may stop using the service and join another service provider (i.e. content provider). Because of the SLA negotiations between the CDN provider and the content provider, the CDN provider will receive penalties on such circumstances. Thus, algorithms need to be aware of QoS metric in order to provide a guaranteed performance. Common QoS metrics include latency and bandwidth.

*3) Reconfigurability:* The configuration of the infrastructure consisting of replica servers does not remain optimal in the long-term, although it may offer cost efficiency and better performance in the short-period that follows the placement. This is because of many factors, such as, fluctuating end-user workload, changing resource prices, etc. A typical reconfiguration job involves addition of new replica servers/resources and deletion of the pre-existing replica servers/resources from the infrastructure. Reconfiguration also involves of switching mode of the replica servers if multi-modal processing is allowed. Reconfiguration can be triggered by phenomena, such as variation in end-user load and variation in QoS performance. In contrast to traditional CDN architectures, cloud based CDN and NFV based CDN architectures can support dynamic reconfiguration because of their ability to provision resources on-demand.

*4) Efficient Resource Utilization:* Balancing the load among replica servers is an important requirement in CDN. Resource utilization is not effective if the entire end-user load is assigned to the closest replica server. Splitting the end-user load among replica server, however, improves the resource utilization.

*5) Demand Resiliency: Resiliency to* demand uncertainty serves an important role in traditional CDNs as resources

cannot be allocated on the fly once a massive surge is detected. Thus, if the placement is decided by considering future uncertainty in traffic, the unexpected upsurge can be managed efficiently. Even for cloud based CDN and NFV based CDN it takes some time to start the VM. It is desirable for a replica server placement algorithm to possess the ability to deal with the uncertainty in end-user demands. The algorithm must incorporate probabilistic or stochastic model to make the placement resilient to demand uncertainty.

6) *Placement Granularity:* It indicates the granularity of candidate replica server locations. Without loss of generality,

it can be assumed that candidate sites for placing servers are large infrastructures having servers of heterogeneous cost and capabilities. It is therefore desirable to select the appropriate number of replica servers per site or allocate the right amount of resources to replica servers at a given site, leading to higher granularity of replica server placement. Generally, replica server placement algorithms must determine the number/size of servers/resources/VNFs in each site in addition to determining the appropriate sites. The placement granularity is low when only one server/VNF is placed per site.

TABLE VIII
COMPARISON OF REPLICA SERVER PLACEMENT ALGORITHMS IN TRADITIONAL CDN
AND EMERGING PARADIGM BASED CDN

| Paper | Cost Minimization | | | Bounded QoS | Reconfigur-ability | Efficient Resource Utilization | Demand Resilience | Placement Granularity |
|---|---|---|---|---|---|---|---|---|
| | Deployment | Delivery | Update | | | | | |
| Algorithms for Traditional CDN | | | | | | | | |
| Jamin *et al.* [24] | N | N | N | N | N | N | N | Low |
| Qiu *et al.* [61] | N | N | N | N | N | N | N | Low |
| Wu *et al.* [44] | N | N | N | N | N | N | N | Low |
| Bhulai *et al.* [43] | N | Y | N | N | N | N | N | Low |
| Hei *et al.* [45] | Y | Y | Implicit | Y (Bandwidth) | N | N | N | Low |
| Sung *et al.* [46] | N | N | Y | Y (Latency) | N | N | N | Low |
| Ho *et al.* [86] | N | Y | N | N | N | N | Y | Low |
| Szymaniak *et al.* [25] | N | N | N | N | N | N | N | Low |
| Yin *et al.* [26] | Y | N | N | N | N | N | N | Low |
| Rodolakis *et al.* [27] | Y | Y | N | Y (Latency) | N | Y | N | Low |
| Kalpakis *et al.* [55] | Y | Y | N | N | N | N | N | Low |
| Goldschmidt *et al.* [47] | N | N | N | N | Y | N | N | Low |
| Xu *et al.* [42] | N | Y | N | N | N | N | N | Low |
| Ahuja *et al.* [48] | N | N | N | Y (Latency) | N | N | N | Low |
| Nguyen *et al.* [54] | Y | Y | N | Y (Distance) | N | Y | Y | Low |
| Chen *et al.* [33] | N | N | N | Y (Latency) | Y | N | N | Low |
| Laoutaris *et al.* [58] | Y | Y | N | N | N | N | N | Low |
| Tang and Xu [70] | Y | N | Y | Y (Distance) | N | N | N | Low |
| Wang *et al.* [69] | N | N | Y | Yes (Distance) | N | N | N | Low |
| Bassali *et al.* [81] | N | Y | N | N | N | N | N | Low |
| Varadarajan *et al.* [82] | N | N | N | N | N | N | N | Low |
| Aupya *et al.* [78] | Y | N | N | N | Y | Y | N | Low |
| Mandal *et al.* [34] | N | N | N | Y (Distance) | Y | N | N | Low |
| Mandal *et al.* [80] | N | N | N | Y (Distance) | Y | N | N | Low |
| Tang *et al.* [72] | N | Y | Y | N | N | N | N | Low |
| Thouin *et al.* [83] | N | Y | N | N | N | Y | N | Low |
| Lim *et al.* [74] | Y | Y | Y | N | N | N | N | Low |
| Bekats *et al.* [99] | Y | Y | N | N | N | N | N | Low |
| Bekats *et al.* [76] | Y | Y | Y | N | N | N | N | Low |
| Luss *et al.* [84] | Y | Y | N | N | N | N | N | Low |
| Benoit *et al.* [56] | Y | N | N | Y (Res. time) | N | Y | N | Low |
| Yang *et al.* [75] | N | Y | Y | N | N | N | N | Low |
| Radoslavov *et al.* [53] | N | N | N | N | N | N | N | Low |
| Laoutaris *et al.* [85] | Y | Y | N | N | N | N | N | Low |

| Algorithms for Cloud based CDN | | | | | | | | |
|---|---|---|---|---|---|---|---|---|
| Rappaport and Raz [28] | Y | Y | Y | N | N | N | N | Low |
| Wang et al. [92] | Y | Y | N | N | N | N | Y | Low |
| Chen et al. [29] | Y | Y | Y | Y (Latency) | N | N | N | Low |
| Papagianni et al. [30] | Y | Y | Y | Y (Latency) | N | N | N | Low |
| Aggarwal et al. [87] | N | N | N | Y (Latency) | N | N | Y | High |
| He et al. [93] | N | Y | N | N | Y | N | Y | High |
| Zhang et al. [88] | Y | N | N | Y (Latency) | Y | N | Y | High |
| Zhenghuan et al. [89] | Y | N | N | Y (Overload prob.) | N | N | Y | High |
| Tran et al. [94] | Y | Y | N | N | Y | N | Y | Low |
| Mokhtarian et al. [91] | Y | Y | Y | N | Y | N | N | High |
| Ferdousi et al. [90] | N | N | N | Y(Latency) | Y | N | N | Low |
| Algorithms for NFV based CDN | | | | | | | | |
| Llorcia et al. [32] | Y | Y | N/A | Y (Latency) | N | N | N | High |
| Bouten et al. [41] | Y | Y | N/A | Y (Latency) | N | N | N | Low |
| Mangili et al. [96] | Y | Y | N/A | Y (Latency) | N | N | Y | High |
| Cohen et al. [31] | Y | Y | N/A | N | N | N | N | High |
| Gupta et al. [98] | N | N | N/A | N | N | N | N | Low |

## B. Observation

TABLE VIII shows the comparison of replica server placement algorithms for traditional CDNs and emerging paradigm based CDNs.

For cost minimization, we observe that deployment cost and delivery cost are used more often than update cost. Although, some algorithms consider either deployment cost or delivery cost, their overall usage are at par. An important remark is that algorithms that overlook delivery cost cannot achieve much economic benefits. This is because, deployment cost is a onetime cost, whereas, delivery cost is a recurring cost, incurred every time the content is delivered to the end-user. As the number of end-users grow, the delivery cost has a major share in the total expenditure.

In terms of bounded QoS, latency is used the maximum number of times followed by geographic distance, bandwidth and response time. Note that, geographic distance is used to represent latency as obtaining real latency information is not very straightforward and requires active measurements or negotiation with a third party to obtain the logs. However, algorithms that use latency as QoS constraint can meet the end-user expectation more easily than with geographic distance. Hei et al. [45] is the only work that uses bandwidth as the QoS indicator.

Apart from Goldschmidt et al. [47] and Aupya et al. [78], all other works lack the ability to reconfigure. In terms of effective resource utilization, only 10% of the algorithms incorporate load balancing. It is mostly because majority of the algorithms adopt the closest server policy and hence the end-users load is not split among the replica servers. Apart from Ho et al. [86], Nguyen et al. [54] all other algorithms for traditional CDNs lack the resiliency against demand uncertainty, leading to downgraded QoS during sudden spike in demand. In terms of placement granularity, none of the algorithms has high granularity i.e. all algorithms place one replica server per candidate site.

Nguyen et al. [54] satisfies the maximum number of requirements (5 out of 10) followed by Rodolakis et al. [27] (4 out of 10). Both Nguyen et al. [54] and Rodolakis et al. meet deployment cost minimization, delivery cost minimization, bounded QoS and efficient resource utilization. Reconfigurability is the requirement satisfaction of which makes Nguyen et al. [54] more suitable than Rodolakis et al. [27].

We observe that a reasonable proportion (36%) of the algorithms for cloud based CDNs minimize all the cost components. Aggarwal et al. [87], however, does not minimize any of the cost, hence lacks cost efficiency. Regarding QoS, latency is the predominant QoS metric used by four algorithms out of five that provide QoS bound. The other metric is overload probability, which is relevant for video based applications. Reconfigurability is satisfied by 40% of the algorithms. It shows that the dynamic provisioning ability of cloud computing paradigm has not been harnessed effectively by the algorithms for cloud based CDN. Interestingly, around 70% of the algorithms satisfy demand resiliency. Despite the relevance of resource utillization in cloud based CDN, none of the algorithms satisfy this requirement. In terms of placement granularity, 50% of the algorithms have high granularity.

In NFV based CDN, all algorithms minimize deployment cost and delivery cost. However, none of them minimize update cost. Surprisingly, reconfigurability is not satisfied by any of the algorithms. Like cloud based CDN, the dynamic provisioning ability of NFV paradigm is not exploited. None of the algorithms meet effective resource utilization. Demand resiliency is met by only one work, that is, Mangili et al. [96], which uses stochastic programming to combat the uncertainty in end-user demands. As far as QoS is concerned, latency is the only QoS metric found and the QoS bound requirement is satisfied by three out of four algorithms. With regard to placement granularity, all but Bouten *et al.* [41] are able to place VNFs with high granularity.

We observe that among all algorithms, algorithms for cloud based CDN and NFV based CDN meet higher number of requirements than the algorithms in traditional CDNs. However the number of traditional CDN algorithms that satisfy the sixth requirement i.e. efficient resource utilization is found to be more than the number of algorithms in cloud based CDN and NFV based CDN that satisfy the same.

## VIII.   FUTURE RESEARCH DIRECTIONS

### A.   Load Balanced Placement

Load balanced placement is necessary in order to ensure effective utilization of replica server resources. If a selected replica server can serve a very small fraction of the end-users while still having capacity, it results in underutilization of resources. As we see in Section.VII, very few works address this issue. Zeng *et al.* [100] introduces a constraint that ensures load balanced placement in cloud based storage systems. The constraint indicates that the potential volume of load for the selected replica server must not be less than the average load over the replica servers that have been selected already. Load balanced placement can also be achieved by minimizing the variance of replica server load.

### B.   Chained VNF Placement in CDN

Existing works on VNF placement in CDN ignore many important aspects, such as, the delay induced by chaining VNFs processing time of VNFs and bandwidth requirements of VNFs. As we have described in Section II-B, VNFs can be chained on the fly to create new value added services in an agile way. Thus, the placement algorithm must place VNFs on the NFVI sites by taking into account the delay between consecutive VNFs of a VNF chain. Processing time is also important as it will impact the delay in delivering the final processed content to the end-users. In addition, the bandwidth requirement at the input and output port of VNFs are different as some VNFs compress/decompress the content according to the underlying functionality. It is therefore imperative to design new algorithms incorporating the above issues for VNF placement in NFV based CDN.

### C.   End-user Demand Prediction and Uncertainty Mitigation

Accurate prediction of end-user demands is critical in placing replica servers. This is because, inaccuracies lead to under or over-provisioning of replica servers/resources. Time series based prediction methods have gained popularity because of their simplicity and effectiveness. They range from the Simple Moving Average (SMA) model to Autoregressive Integrated Moving Average (ARIMA) model [101]. Principal component analysis (PCA) [102] is another effective way to forecast demand. In [103], PCA is used to obtain video demand evolution patterns over a longer time intervals (e.g.

weeks or months). Time series analysis based models are used in some algorithms for cloud based CDN [92]. However, none of the algorithms in NFV based CDN contains a forecast component and hence the above methods can be investigated to show the benefits of demand prediction on cost and performance of a placement.

Prediction models are suitable for leasing resources over a short period. However, for long leasing periods, uncertainty in end-user demand/workload needs to be considered in replica server placement. High uncertainty increases the possibility of economic risk for the CDN provider. At the same time, the advantage of considering workload uncertainty in replica server placement is that placement obtained for a random scenario fits well in all scenarios. Thus, it is necessary to consider a random work-load. In the literature, uncertainty has been addressed by using optimization frameworks, such as, robust optimization [86], and stochastic programming [96]. Robust optimization is used when the workload is given as a range of values, whereas, in stochastic optimization the workload is represented as a probability distribution. Other techniques such as Stochastic Dynamic Programming (SDP) can be investigated for mitigating uncertainty. Simple methods, such as mean-traffic model [86] and worst-case traffic model [86] can also be used as benchmarks to assess the efficiency of a proposed replica server placement algorithm that considers uncertainty.

### D.   Transmission Link Cost Optimization

In most of the cloud based CDNs discussed in the paper, the replica servers are built by leasing resources from one or more cloud providers. In other words, CDN provider and cloud provider are two different business entities. Because of resource leasing, the delivery cost denotes the bandwidth leasing cost for uploading and downloading content to and from the replica servers. However, some commercial cloud based CDN (e.g. Amazon CloudFront) is operated by the same entity i.e. the cloud provider. Thus, these CDNs incur higher delivery cost because of the high capacity transmission link that connects the cloud sites. In order to achieve cost-efficiency, the cost of the transmission links needs to be minimized. It is therefore imperative to formulate new optimization models by incorporating the above cost component in the objective function and suitable heuristics need to be designed.

### E.   Metaheuristics

Current scenario of replica server placement involves large problem instances. For instance, when a cloud based CDN architecture is built by relying on multiple cloud providers, the number of candidate sites can be up to 1000. At the same time, the number of end-users would be in the order of millions. Needless to say that metaheuristics provide good enough solutions for large-size problem instances and hence need to be investigated for cloud based CDN.

The motivation for using metaheuristics is even more prominent in NFV based CDN. This is because, VNF placement is more complex than replica server placement. In particular, each end-user requests a certain VNF chain consisting of a set of VNFs with a performance guarantee, which can be a threshold on delay. Due to heterogeneous infrastructure resources and capacities, the VNFs of a chain are placed on a NFVI site that fulfills the requirements of VNFs. Thus, VNF placement has one more dimension, that is, the number of VNFs that attributes to the size of input and impacts the execution time of the algorithm. For example, the greedy based algorithms presented in [104] have high time complexity. Considering the huge size of the VNF placement problem, it is worth investigating the use of metaheuristics.

From the review, we found that very few (only three) algorithms are based on meta-heuristics. In traditional CDN, Wu *et al.* [44] investigated GA for replica server placement, wherein the only criterion is the hop-count of the shortest path between the end-users and the replica servers. In cloud based CDNs, metaheuristics are not investigated at all. In NFV based CDN, Bouten *et al.* [41] is the only work on VNF placement that uses GA.

*F. Multi-objective Framework*

The multitude of requirements we have identified in Section-VII. A dictate the need to design a holistic multi-objective optimization framework to formulate a replica server placement problem. The potential objectives include cost minimization and efficient resource utilization. The simplest way to solve multi-objective optimization is to adopt the weighted sum approach in which a single objective function is created by linearly combining all objectives. Another approach involves optimizing only one of the objectives and representing others as constraints. In contrast to above approaches, pareto-optimal optimization provides the flexibility to analyze the trade-offs among objectives and hence is worth investigating. In Yin *et al.* [26], the media server placement is solved using pareto-optimal optimization. But, cost components, such as delivery cost and update cost are ignored. Moreover, the algorithm does not ensure effective resource utilization.

*G. Server Placement Algorithms in Cloudlet based CDN*

Cloudlet [105] [106]is another emerging paradigm introduced to supplement cloud computing paradigm for reducing the delay of real-time services offered to mobile end-users, especially those which provide cognitive assistance, such as, augmented reality, speech recognition, navigation, language translation ( [107], [105]) provided to end-users. Cloudlets are clusters of computers collocated at access points. The close proximity of these tiny datacenters with the end-users results in significant delay reductions compared to remote cloud, thereby making them ideal candidates to host the replica servers in CDN and to offer value added services to the end-users. For example, in case of CDN that hosts dynamic content, such as social media, the weather update can be delivered to the end-user from a cloudlet. Akamai has started offering cloudlet based value added services (e.g. Image Converter) to allow end-users to dynamically manipulate images [108].

In a typical cloudlet approach based on public cloud, the CDN provider leases resources from the cloudlet provider

(e.g. local government) and places the servers on suitable cloudlets. To design, efficient algorithms are needed to place CDN servers on the available cloudlet sites. Few works have addressed Cloufdlet placement and assignment of end-users to a suitable cloudlet. In[109], [110] cloudlet placement has been studied for wireless metropolitan area networks (WMAN). The existing cloudlet placement models need to be studied analyzed and can be extended to solve CDN server placement in cloudlet.

## IX. CONCLUSION

In this survey, we provide an in-depth discussion of replica server placement in traditional, cloud based and NFV based CDNs. The summary of various algorithms show that there are eight different objectives (RTT, latency, hop-count, link quality, cost, number of servers, number of reconfigurations, network traffic) proposed in the literature. With regard to the problem formulation, integer programming is the most common optimization framework adopted, while very few of the works adopt stochastic programming. As far as solution approach is concerned, majority (42%) of the algorithms are designed using a greedy or modified greedy approach. Metaheuristics are used by only 6% of the algorithms. Approaches such as dynamic programming, Nash bargaining solution and model predictive control are also investigated.

After elucidating the characteristics of the algorithms, we identified requirements for an efficient replica server placement algorithm and reviewed the existing algorithms in light of these requirements. The most interesting observation from the comparison is that the algorithms for cloud based CDN and NFV based CDN meet the maximum number of requirements than the algorithms for traditional CDNs.

We conclude by a discussion on the potential avenues for future research in replica server placement. Some of them include the need for accurate prediction schemes, investigation of metaheuristics and design of replica server placement for CDNs using other emerging paradigm such as Cloudlet.

## ACKNOWLEDGEMENT

This work was supported in part by Ericsson Canada, and the Natural Science and Engineering Council of Canada (NSERC) through the SAVI Research Network. We would also like to thank Elaheh T. Jahromi for her contributions to the use case described in the paper.

**Jagruti Sahoo** received a Ph.D. degree in computer science and information engineering from the National Central University, Taiwan, in January 2013. She worked as Postdoctoral Fellow in University of Sherbrooke, Canada and Concordia University, Canada. She is currently an Assistant Professor at the Department of Mathematics and Computer Science, South Carolina State University, USA. Her research interests include wireless sensor networks, vehicular networks, content delivery networks, cloud computing and network functions virtualizations. She has served as a member of the technical program committee of many conferences and as a reviewer for many journals and conferences.

**Mohammad A. Salahuddin** is a Postdoctoral Fellow with the Department of Computer Science at Université du Québec à Montréal, Montreal, Canada. He is also a Visiting Scientist with the Concordia Institute for Information Systems Engineering at Concordia University, Montreal, Canada. He received his Ph.D. degree in Computer Science from Western Michigan University, Kalamazoo, Michigan, USA, in 2014. His research interests include Wireless Sensor Networks, QoS and QoE in Vehicular Ad hoc Networks (WAVE, IEEE 802.11p and IEEE 1609.4), Internet of Things, Content Delivery Networks, Software-Defined Networking, NFV and Cloud Resource Management. He serves as a Technical Program Committee member and reviewer for IEEE journals, magazines and conferences.

**Roch Glitho** holds a Ph.D. (Tekn.Dr.) in tele-informatics (Royal Institute of Technology, Stockholm, Sweden) and M.Sc. degrees in business economics (University of Grenoble, France), pure mathematics (University Geneva, Switzerland), and computer science (University of Geneva). He works in Montreal, Canada, as an Associate Professor of networking and telecommunications at the CIISE where he leads the telecommunication service engineering research laboratory (http://users.encs.concordia.ca/~tse/). In the past he has worked in industry for almost a quarter of a century and has held several senior technical positions at LM Ericsson in Sweden and Canada (e.g. expert, principal engineer, senior specialist). His industrial experience includes research, international standards setting (e.g. contributions to ITU-T, ETSI, TMF, ANSI, TIA, and 3GPP), product management, project management, systems engineering and software/firmware design. In the past he has served as IEEE Communications Society distinguished lecturer, EditorIn-Chief of IEEE Communications Magazine and Editor- In-Chief of IEEE Communications Surveys & Tutorials.

**Halima Elbiaze** holds a Ph.D. in computer science and a M.Sc in Telecommunication systems from Institut National des Télécommunications, Paris, France and Université de Versailles in 2002 and 1998, and B.S. Degree in applied mathematics from university of MV, Morocco in 1996. Since 2003, she is with the Department of Computer Science, Université du Québec à Montréal, QC, Canada, where she is currently an Associate Professor. She is the author or coauthor of many journal and conference papers. Her research interests include network performance evaluation, traffic engineering, quality of service management in optical and wireless networks.

**Wessam Ajib** received an Engineer diploma from INPG at Grenoble, France in Physical Instruments, in 1996, a Master and Ph.D. degrees in computer networks from École Nationale Supérieure des Télécommunication, Paris, in 1997 and in 2000. He had been an architect and radio network designer at Nortel Networks, Ottawa, ON, Canada between October 2000 and June 2004. He had conducted many projects on the third generation of wireless cellular networks. He followed a post-doc fellowship at Electrical Engineering department of École Polytechnique de Montréal, QC, Canada in 2004-2005. Since June 2005, he has been with the Department of Computer Sciences, Universite du Quebec at Montreal, QC, Canada, where he is presently a full Professor. His research interests include wireless communications and networks, multiple access design, and traffic scheduling. He is the author or co-author of many journal papers and conferences papers in these areas.